\documentclass[ preprint,11pt]{elsarticle}
\usepackage[a4paper, total={6.5in, 10in}]{geometry}
\usepackage{setspace}
\usepackage{ amsmath,  amssymb }
\usepackage{amsthm}
\usepackage[para,online,flushleft]{threeparttable}

\usepackage[load=prefixed, load=abbr]{siunitx}
\usepackage{enumerate}
\biboptions{sort&compress}
 \usepackage{caption} \usepackage{array,booktabs}
\usepackage{graphicx} 
\usepackage{multirow} 
\usepackage{subcaption}
\usepackage{hyperref}

\DeclareMathOperator*{\argmax}{arg\,max}
\DeclareMathOperator*{\minimize}{minimize}

\theoremstyle{remark} \newtheorem{remark}{Remark}

\journal{Mechanical Systems and Signal Processing}

\begin{document}

\begin{frontmatter}


\title{ Primary-Auxiliary Model Scheduling Based Estimation of  the Vertical Wheel Force  in   a Full Vehicle System}

\affiliation[inst1]{organization={University of Michigan -- Shanghai Jiao Tong University Joint Institute, Shanghai Jiao Tong University},
            addressline={800 Road, Minghang district}, 
            city={Shanghai},
            postcode={200240}, 
            country={China}}

\affiliation[inst2]{organization={School of Mechanical Engineering, Shanghai Jiao Tong University},
            addressline={800 Road, Minghang district}, 
            city={Shanghai},
            postcode={200240}, 
            country={China}}

\affiliation[inst3]{organization={SAIC Motor Corporation Limited},
            addressline={201 Anyan Road, Jiading district}, 
            city={Shanghai},
            postcode={201800}, 
            country={China}}

\affiliation[inst4]{organization={Department of Automation, Shanghai Jiao Tong University},
            addressline={800 Road, Minghang district}, 
            city={Shanghai},
            postcode={200240}, 
            country={China}}

\author[inst1]{Xueke Zheng}

\author[inst1,inst2]{Runze Cai\corref{cor1}}
\ead{cairunze@sjtu.edu.cn}
\author[inst1]{Shuixin Xiao}

\author[inst3]{Yu Qiu}

\author[inst1,inst4]{Jun Zhang}

\author[inst1,inst4]{Mian Li}
\cortext[cor1]{Corresponding author.}

\begin{highlights}
\item A modeling scheduling   framework for sensor-to-sensor identification problems for the purpose of signal estimation is proposed.
\item The method is  applicable to the systems with switching linear dynamics.
\item An experimental validation on a full vehicle system  is performed to reconstruct  the vertical wheel force.
\item Good accordance between the measurements and the estimates  is achieved  with the reasonable accuracy level under multiple working conditions.

\end{highlights}

\begin{abstract}
  In this work we study    estimation problems  in nonlinear mechanical  systems subject to non-stationary and unknown excitation, which are common and critical problems in design and  health management of mechanical  systems.

  A primary-auxiliary model scheduling procedure based on time-domain transmissibilities is proposed and performed  under switching linear dynamics:  In addition to constructing a primary transmissibility family from the pseudo-inputs to the output during the offline stage, an auxiliary transmissibility family is constructed by further decomposing the pseudo-input vector into two parts. The auxiliary family enables to determine   the unknown working condition at which  the system is currently running at, and then   an appropriate transmissibility from the primary transmissibility family for estimating the unknown output can be selected  during the online estimation stage. Moreover, Finite Impulse Response (FIR) models, ridge regression, and Bayes classifiers are applied to realize the model scheduling procedure. As a result, the proposed approach     offers a   generalizable and explainable    solution to the   signal estimation problems in nonlinear mechanical systems  in the context of  switching linear dynamics  with unknown  inputs.

    A numerical example  and a real-world application to  the estimation of the vertical wheel force  in a  full vehicle system   are, respectively,  conducted to demonstrate the effectiveness of the proposed method.  During the vehicle design phase, the vertical wheel force  is the most important one among   Wheel Center Loads (WCLs),   and  it is  often measured directly with expensive, intrusive, and hard-to-install measurement devices during full vehicle testing campaigns.  Meanwhile, the  estimation problem of the vertical wheel force    has not been solved well  and is still of great  interest.  The experimental results  show good performances of the proposed method in the sense of  estimation accuracy for estimating the vertical wheel  force.
\end{abstract}


\end{frontmatter}

\section{Introduction}

When designing a new vehicle, the durability performances can be assessed by considering the  knowledge of Wheel Center Loads (WCLs), i.e., longitudinal, lateral, and vertical wheel forces, camber, torque, and steer moments,  as  input quantities~\cite{johannesson2013guide}. In the current practice, the  WCLs are measured directly by so-called Wheel Force Transducers (WFTs) during Road Load Data Acquisition (RLDA) testing campaigns in which  prototype vehicles are  driven on proving grounds or public roads. However, WFTs are expensive, intrusive, and time-consuming to install. Particularly, it is economically infeasible to install  WFTs in each vehicle when multiple vehicles  need to be tested~\cite{elkafafymachine, risaliti2019multibody}.  Therefore, in order  to facilitate the vehicle design phase, it is appealing to resort to alternative ways of obtaining WCLs without directly measuring them by WFTs. 

\subsection{Physically  Based  and Data-driven Methods}
\label{sec:phys-based-data}
The vehicle dynamic system  is a  complex and nonlinear mechanical system~\cite{risaliti2019multibody, risaliti2018virtual, elkafafymachine}.  In order to accurately estimate  WCLs by some other easy-to-measure quantities,  E. Risaliti et al.~\cite{risaliti2019multibody,risaliti2018virtual} develop an approach  based on  multi-body simulation models and  augmented Extended Kalman Filters (EKFs). The multi-body based approach achieves a good performance in a McPherson suspension system  in~$4$ experimental runs. However,  the  approach may have  three disadvantages: First, the highly accurate multi-body simulation model is  not easy  to obtain in some cases~\cite{elkafafymachine}; Second,  the model  parameters  of EKFs are difficult to determine   when  prototype vehicles are driven under a variety of  working conditions  on   proving grounds or  public roads (in which cases the excitation of the system is unknown and non-stationary); Finally,  the use of EKFs  might be problematic when the system is  strongly nonlinear~\cite{risaliti2019multibody}. Nevertheless, the main advantage of using multi-body based approach is to gain more physical insights on the effects of the variation of the system parameters~\cite{elkafafymachine}.

An alternative is to use data-driven models instead of multi-body simulation models to build the input-output relations between WCLs and other measurements such as the strain gauges,  suspension deflections and accelerations. Several common data-driven models such as linear Auto-Regressive with eXtra input (ARX),  Polynomial Nonlinear State Space (PNLSS), and the Recurrent Neural Network (RNN) models are  used to estimate WCLs. However, it turns out that the linear ARX models cannot capture the nonlinear effects,  and PNLSS as well as  RNN models lack the generalization ability~\cite{elkafafymachine}. In addition,  the computational cost and memory requirements are challenging factors for real-time systems when the nonlinear models  are involved. 
As a result, the estimation problem of WCLs in  RLDA testing campaigns has not been solved well  and is   of great  interest, hence  a simple and  effective  method is in high demand.

It is worthwhile to notice that  the vertical wheel force  not only is the most important load among WCLs, but also is most affected by the variation of model parameters in multi-body model based method~\cite{risaliti2019multibody}. Therefore, a successful estimation of the vertical wheel force is a good  step  for  the estimation of WCLs.

\subsection{Sensor-to-Sensor  and Piece-Wise Affine (PWA) System Identification}
\label{sec:sensor-sensor-piece}
Sensor-to-sensor system identification problems  have received  considerable research interests in recent past~\cite{yan2019transmissibility,aljanaideh2018experimental, d2009sensor,brzezinski2011identification, aljanaideh2015time,linder2017identification}. In this type of problems, the relation between the response of a subset of sensors   and the response of the remaining ones is modeled in the frequency domain or in the  time domain~\cite{devriendt2010operational, chesne2013damage, yan2019transmissibility, RIBEIRO200029, aljanaideh2015time}, and such type of modeling is extensively required by the applications of structural modeling and structure health monitoring systems~\cite{OMA2011,WEIJTJENS2014559,kukreja2012sensor,feng2015damage}. In this area, it is often assumed that the system is linear~\cite{brzezinski2010sensor,brzezinski2011identification,aljanaideh2015time}, and transmissibilities are introduced to model the relation between the signals from sensors without knowing the excitation of the system, which can be  non-stationary. 
Moreover, due to  good properties   such as Bounded-Input Bounded-Output (BIBO) stability, Finite Impulse Response (FIR)  models  have  proven to be better candidates over ARX models for  estimating  transmissibilities   in many sensor-to-sensor identification problems~\cite{aljanaideh2018experimental, aljanaideh2015time2, ALJANAIDEH2020108686}. However, FIR models  are also linear  and thus  directly  applying them to the nonlinear mechanical systems  will lead to poor performances (which are also shown in our experimental results in Section~\ref{sec:appl-vehicle-health}).

Recent years have witnessed a growing interest on Piece-Wise Affine (PWA) system identification methods which have proven to be effective for problems involving complex nonlinear systems with large data sets~\cite{garulli2012survey, paoletti2007identification}. Many literatures focus on studying piecewise ARX  models for the purpose of system control~\cite{garulli2012survey,paoletti2007identification}. However, when it comes to classifying the regressor domain of piecewise FIR models for the purpose of signal estimation, data classification should be carefully addressed since the dimension of the regressor domain is often much higher.

\subsection{The Proposed Method: Primary-Auxiliary  Model Scheduling Procedure}
\label{sec:prop-meth-model}
In this work, we consider the situation that the data of the vertical
wheel force  and some other easy-to-measure quantities 
(e.g., accelerations  and suspension deflections) obtained by
affordable and common sensors   is collected from multiple working conditions on  proving grounds, and our goal is to use these   easy-to-measure quantities   to estimate the vertical wheel force  in a full  test vehicle.

 We validate that a   full vehicle dynamic system  can be approximated by a switching linear dynamic system~\cite{liberzon2003switching} with unknown inputs, and   each sub-system describes the dynamics of the system running at an unknown  working condition. Note that   determining the unknown working conditions during the online estimation stage  is the bottleneck in the estimation of the vertical wheel force.

The response of easy-to-measure quantities   is  essentially the outputs of the true system in the  sensor-to-sensor system identification context. Based on this observation, besides constructing a primary time-domain transmissibility family from  the response of easy-to-measure quantities to the  vertical wheel force  under multiple working conditions during the offline stage, another auxiliary time-domain transmissibility family is constructed by decomposing  these easy-to-measure quantities into two parts. The auxiliary transmissibility family   is further used for constructing the Bayes classifier which determines the unknown working conditions during the online estimation stage, and  then an appropriate transmissibility from the primary transmissibility family is selected for estimating the unknown  vertical wheel force.

\subsection{Main Contributions and Paper Outline }
\label{sec:main-contr-paper}
The main contribution of this paper includes: We propose a primary-auxiliary  model scheduling procedure  for signal estimation  in nonlinear mechanical systems subject to non-stationary and unknown excitation, which  can be approximated by switching linear systems with unknown inputs.   The key reason behind the success of the proposed method is the strategic construction of the auxiliary transmissibility family which enables model scheduling during the online estimation stage. Thus, the unknown working conditions are determined in a sensible manner.  The benefits of introducing the auxiliary transmissibility family are
\begin{itemize}
\item[$\bullet$]
The construction of FIR models and the Bayes classifier is computationally tractable; no nonlinear or nonconvex optimizations are involved.
\item[$\bullet$]
The transmissibility family allows to avoid classifying the high dimensional regressor, which could lead to poor performances in PWA system identification methods. 
\end{itemize}
For the reason above, the proposed method    offers a   generalizable and explainable    solution to the   signal estimation problems in nonlinear mechanical systems  in the context of  switching linear dynamics  with unknown  inputs. 

 A real experiment representing an industrial application (i.e., the estimation of the vertical wheel force  in RLDA testing campaigns)  as a particular case   is used to validate the proposed  methodology. In comparison with  limited experimental runs  carried out with  a   suspension system in work~\cite{risaliti2019multibody, risaliti2018virtual, elkafafymachine},  the experimental runs are carried out under a variety of working conditions (covering the typical ones in RLDA testing campaigns  as many as possible) with  a full vehicle system in this work, which leads to more realistic  experimental settings  for RLDA testing campaigns. Since the six WCLs can be estimated independently within the framework of the primary-auxiliary model scheduling, the estimation of the vertical wheel force can be extended to the estimation of the  other five WCLs.

The rest of the paper is organized as follows: In Section~\ref{sec:FIR model}, we briefly introduce FIR models, Maximum Likelihood Estimation (MLE), and ridge regression. In Section~\ref{sec:patt-recogn-based}, a primary-auxiliary model scheduling procedure during  the online estimation stage  is detailed. In Section~\ref{sec:numerical-example}, a numerical example of a quarter-car suspension system  is shown, and  a real-world application  to  the vertical wheel force  estimation in a full vehicle system  is demonstrated for the effectiveness of the proposed method   in Section~\ref{sec:appl-vehicle-health}. Finally, conclusion remarks and future works  are presented in Section~\ref{sec:conclusion}.

\section{Linear Regression for FIR Models}
\label{sec:FIR model}
\begin{figure}[t!]
\centering
\includegraphics[page=1,width=3.25in]{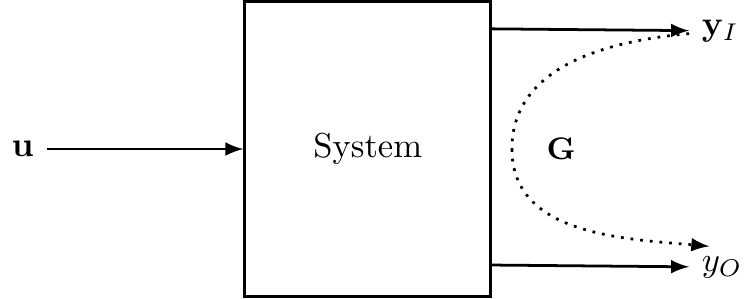}
\caption{The transmissibility from pseudo-inputs $\mathbf{y}_I$ to output $y_O$.}
\label{fig:S2SID1}
\end{figure}

Consider a dynamical system as shown in Fig.~\ref{fig:S2SID1}, where~$\mathbf{u}$ denotes the excitation or the inputs to the system, and $\mathbf{y}_I$ as well as $y_O$ are the outputs of the system. In this section, we assume that the system is linear, and the relation from $\mathbf{y}_I$ to $y_O$ can be described as a discrete-time transmissibility, denoted by $\mathbf{G}$, which will be used later  in the nonlinear system context in Section~\ref{sec:patt-recogn-based}.

\subsection{Maximum Likelihood Estimation (MLE) of FIR Model Parameters}
\label{subsec: MLEFIR}

Now consider the following Multiple-Input Single-Output (MISO) model for $\mathbf{G}$ given by

\begin{equation}
\label{FIR model}
y_O(t)=\mathbf{b}_0^T \mathbf{y}_I(t)+\ldots+\mathbf{b}_{n}^T \mathbf{y}_I(t-n)+\epsilon(t),
\end{equation}
where $\mathbf{y}_I(t) \in \mathbb{R}^{n_I}$ and $y_O(t) \in \mathbb{R}$ are, respectively, noise-free pseudo-inputs and output measurements of transmissibilities, $n_{I}$ is the the dimension of the pseudo-inputs, $\mathbf{b}_{i} \in \mathbb{R}^{n_I}$ are parameters, $i=0, \dots, n$, and $\epsilon(t)$ is the residual representing the unmodeled dynamics. We intend to identify the parameters $\mathbf{b}_{i}$'s given $\mathbf{y}_{I}$ and~$y_{O}$ over a period of time, say, $t=n+1, \dots, N+n$. 
Now it is easily verified that Eq.~\eqref{FIR model} can be reformulated as a linear regression problem
\begin{equation}
\label{linear regression}
y_O(t) = \mathbf{\phi}_I^T(t) \mathbf{\theta} +\epsilon(t), \qquad t=n+1,\dots,N+n, 
\end{equation}
where $\mathbf{\theta}=
\bigl[\mathbf{b}_0^T  \ \cdots \  \mathbf{b}_{n}^T \bigr]^T $, and we assume that $\epsilon(t)$ is an i.i.d.~Gaussian random variable for all~$t$, and the regressor $\mathbf{\phi}_{I}(t)= \bigl[  \mathbf{y}_I^T(t) \ \cdots \ \mathbf{y}_I^T(t-n) \bigr]^T$. 
Rewriting Eq.~\eqref{FIR model} by stacking the elements (rows) $\mathbf{y}(t)$ and $\mathbf{\phi}_{I}^T(t)$ in the vectors (matrices) yields the vector form of the regression problem as
\begin{equation}
\label{eq:linear_regression_vec}
\mathbf{Y}_O=\mathbf{\Phi}_I\mathbf{\theta}+\mathbf{E}, 
\end{equation}
where
\begin{gather}
\mathbf{Y}_O=
\begin{bmatrix}
y_O(n+1) \\ \vdots \\ y_O(N+n) 
\end{bmatrix},
\mathbf{\Phi}_I=
\begin{bmatrix}
\mathbf{\phi}_I^T(n+1) \\ \vdots \\ \mathbf{\phi}_I^T(N+n)
\end{bmatrix}
, 
\mathbf{E}=\begin{bmatrix}
\epsilon(n+1)\\ \vdots \\ \epsilon(N+n) \end{bmatrix}.
\end{gather}

A statistical argument for estimating $\mathbf{\theta}$ in Eq.~\eqref{eq:linear_regression_vec} is that the measurement $\mathbf{Y}_{O}$ can be regarded as a realization of a random variable with a normal distribution

\begin{equation}
\label{eq:gaussian_dis}
\mathbf{Y}_O \sim \mathcal{N}(\mathbf{\Phi}_I\mathbf{\theta},\sigma^2 \mathbf{I}),
\end{equation}
where $\sigma^2$ is noise variance of $\epsilon(t)$, and $\mathbf{I}$ is an identity matrix of the appropriate dimension.
By solving the optimization problem

\begin{equation}
\label{eq:least-squares}
\begin{aligned}
\minimize_{\mathbf{\theta}} & & \lVert \mathbf{Y}_O -\mathbf{\Phi}_I \mathbf{\theta}\rVert_2^2 
\end{aligned},
\end{equation}
the maximum likelihood estimator of $\mathbf{\theta}$ is given by

\begin{equation}
\label{eq:unregu_MLE_theta}
\hat{\mathbf{\theta}}^{\text{MLE}}=\bigl(\mathbf{\Phi}_I^T\mathbf{\Phi}_I\bigr)^{-1}\mathbf{\Phi}_I^T\mathbf{Y}_O,
\end{equation}
provided that $\mathbf{\Phi}_I^T \mathbf{\Phi}_I$ is non-singular, which is guaranteed if the pseudo-inputs~$\mathbf{y}_I$ is persistently exciting~\cite{lennart1999system}.
\begin{remark}
There are two reasons of modeling the relation between $\mathbf{y}_{I}$ and $y_O$ by an FIR structure: 
\begin{enumerate}[(i)]
\item
An FIR model structure is Bounded-Input Bounded-Output (BIBO) stable even if the transfer function from the excitation to the pseudo-inputs is a non-minimum phase system, i.e., there are zeros outside of the unite circle in the discrete-time case, which indicates that the transmissibility is unstable. It is worth noting that the instability of the transmissibility induces noncausal components in the transmissibility impulse response~\cite{kukreja2012sensor}, in which case a noncausal FIR model can be used to model the relation between the pseudo-inputs and the output with minor changes~\cite{ALJANAIDEH2020108686}.
\item 
An FIR model provides a consistent estimate of true parameters if the sequence $\{\mathbf{y}_I(t)\}$ is independent of the sequence $\{\epsilon(t)\}$. A detailed study on the consistency of least squares methods for modeling parameters in the presence of uncorrelated and correlated input, process, and output noise can be found in~\cite{fledderjohn2010comparison}.
\end{enumerate}
\end{remark}

\subsection{Ridge Regression of FIR Model Parameters}
\label{regu_MLE_FIR}

Sometimes occasions may arise when the pseudo-inputs are poorly exciting during the identification of FIR models. In such cases, one issue of the FIR model is the selection of the model order $n$ which is usually unknown in advance. For the model parameter $\mathbf{\theta}$, an inappropriate selection of $n$ can lead to either a large bias or variance. To strike a balance between the bias and variance, a successful method is ridge regression~\cite{ljung2013can}. Applying ridge regression, the regression problem in Eq.~\eqref{eq:linear_regression_vec} can be solved by considering the regularized least squares problem
\begin{equation}
\label{eq:regu_ls}
\begin{aligned}
\minimize_{\mathbf{\theta}} & & \lVert \mathbf{Y}_O- \mathbf{\Phi}_I \mathbf{\theta} \rVert_2^2+\rho \mathbf{\theta}^T \mathbf{\theta}, 
\end{aligned}
\end{equation}
where $\rho \mathbf{\theta}^T \mathbf{\theta}$ is a flexibility penalty term, and $\rho$ is a tuning parameter~\cite{ljung2013can}. The optimization problem~\eqref{eq:regu_ls} admits a closed-form solution

\begin{equation}
\label{eq:regu_theta}
\hat{\mathbf{\theta}}^{\text{ridge}}= \left(\mathbf{\Phi}_I^T\mathbf{\Phi}_I+\rho \mathbf{I} \right)^{-1}\mathbf{\Phi}_I^T \mathbf{Y}_O.
\end{equation}
Furthermore, the variance $\sigma^2$ can be estimated as~\cite{magnus2019matrix}
\begin{equation}
\label{eq:ridge_variance}
\hat{\sigma}^2=\frac{\lVert \mathbf{Y}_O-\mathbf{\Phi}_I \hat{\mathbf{\theta}}^{\text{ridge}}\rVert_2^2}{N-n_I(n+1)}.
\end{equation}
The ridge regression problem in Eq.~\eqref{eq:regu_ls} is frequently used in statistics to overcome the issue of ill-conditioning in the original linear regression problem. The tuning parameter $\rho$ is usually chosen by cross-validation or other nonlinear optimization methods~\cite{ljung2013can,chen2013implementation}.

In this work we choose $\rho$ in a simpler manner. Let $\lambda_{\max}$ and $\lambda_{\min}$ denote the largest and smallest eigenvalues of $\mathbf{\Phi}_I^T\mathbf{\Phi}_I$, respectively. In order to avoid the matrix $\mathbf{\Phi}_I^T\mathbf{\Phi}_I+\rho \mathbf{I}$ to be close to being singular, the condition number of $\mathbf{\Phi}_I^T\mathbf{\Phi}_I+\rho \mathbf{I}$, denoted by $\kappa(\mathbf{\Phi}_I^T\mathbf{\Phi}_I+\rho \mathbf{I})$, is controlled below a reasonable level denoted by $C_{\text{lim}}$, e.g., $C_{\text{lim}}=\num{1e6}$. It can be easily verified that the largest and smallest eigenvalues of the matrix $\mathbf{\Phi}_I^T\mathbf{\Phi}_I+\rho \mathbf{I}$ are given by $\lambda_{\max}+\rho$ and $\lambda_{\min}+\rho$, respectively. We set the condition number of the matrix $\mathbf{\Phi}_I^T\mathbf{\Phi}_I+\rho \mathbf{I}$ below $C_{\text{lim}}$, i.e.,
\begin{equation}
\label{inequality}
\kappa(\mathbf{\Phi}_I^T\mathbf{\Phi}_I+\rho \mathbf{I})= \frac{\lambda_{\max}+\rho}{\lambda_{\min}+\rho} \leq C_{\text{lim}}.
\end{equation} 
It can be verified that the inequality~\eqref{inequality} is satisfied if we choose $\rho$ as
\begin{equation}
\rho= 
\begin{cases}
0, & \kappa(\mathbf{\Phi}_I^T\mathbf{\Phi}_I) \leq C_{\text{lim}}, \\[2ex]
\dfrac{\lambda_{\max}-\lambda_{\min} \cdot C_{\text{lim}}}{C_{\text{lim}}-1}, & \kappa(\mathbf{\Phi}_I^T\mathbf{\Phi}_I) > C_{\text{lim}}. 
\end{cases}
\end{equation}

\section{Primary-Auxiliary Model Scheduling for Signal Estimation}
\label{sec:patt-recogn-based}

In this section, we consider the system in Fig.~\ref{fig:S2SID1} in a nonlinear setting, and the primary-auxiliary  model scheduling procedure is deployed based on FIR models.

\subsection{Model Scheduling Based on  Primary and Auxiliary Transmissibility Families}
\label{sec:gain_schedule}
\begin{figure}[t!]
\centering
\includegraphics[page=2,width=3.25in]{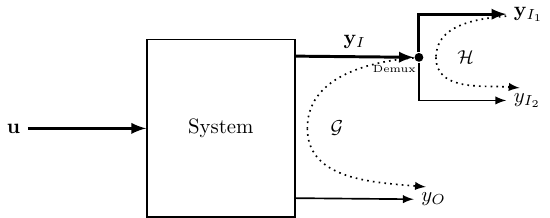}
\caption{transmissibilities from $\mathbf{y}_I$ to $y_O$, and $\mathbf{y}_{I_1}$ to $y_{I_2}$.}
\label{fig:S2SID2}
\end{figure}

As mentioned previously, the performance of modeling the system by one single FIR model is possibly poor due to   non-linearity issues. 
As is commonly described in PWA system identification problems~\cite{garulli2012survey}, suppose the system can be approximately described by multiple transmissibilities, each of which provides a satisfactory description of the system under a different working condition $\mathcal{C}_q$, where $q \in \mathcal{Q}=\{1,\ldots, Q\}$. Explicitly, the relation from $\mathbf{y}_I$ to $y_O$ can be represented by a transmissibility family $\mathcal{G}=\{ \mathbf{G}_q\mid q \in \mathcal{Q}\}$, where $\mathbf{G}_q$'s can be approximately modeled by FIR models introduced in Section~\ref{sec:FIR model}. Then the problem comes down to determining which working condition the system is currently at when estimating $y_O$. In sensor-to-sensor problems  shown in Fig.~\ref{fig:S2SID2}, if the pseudo-inputs $\mathbf{y}_I$ can be decomposed as

\begin{equation}
\label{eq:1}
\mathbf{y}_{I}=
\begin{bmatrix}
\mathbf{y}_{I_1}\\y_{I_{2}}
\end{bmatrix}, 
\end{equation}
where $\mathbf{y}_{I_1}(t) \in \mathbb{R}^{n_{I_1}}$, $y_{I_2}(t) \in \mathbb{R}$ for each $t$, whenever the dimension of $\mathbf{y}_I$ is greater than one (which is the case here for~$\mathbf{G}$), thus the relation from $\mathbf{y}_{I_1}$ to $y_{I_2}$ can be described by another transmissibility family $\mathcal{H}=\{ \mathbf{H}_q\mid q \in \mathcal{Q}\}$ modeled by FIRs. The family $\mathcal{H}$ enables us to connect FIR models with data classification in pattern recognition, where we aim to find a decision rule which assigns a class to each observation based on the system outputs~\cite{bishop2006pattern}.

\begin{figure*}[h!]
\centering
\includegraphics[width=\textwidth]{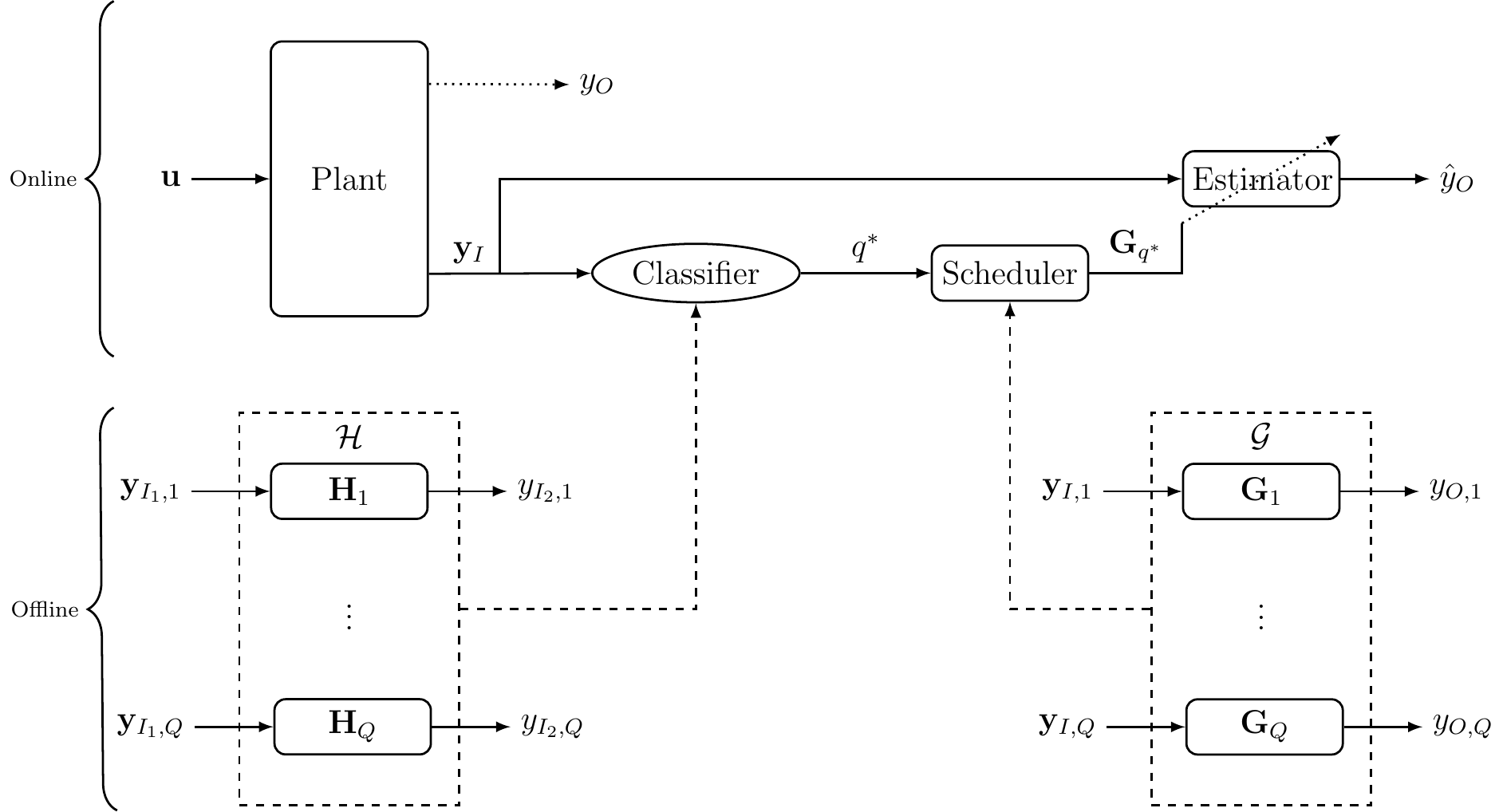}
\caption{The overall procedure of the sensor-to-sensor system identification and primary-auxiliary model scheduling  during offline and online stages, respectively.}
\label{fig:procedure}
\end{figure*}

The overall procedure proposed in this work is shown in Fig.~\ref{fig:procedure}, which consists of two stages: During the online stage, the pseudo-inputs $\mathbf{y}_I$ is known and a Bayes classifier is used to choose the transmissibility $\mathbf{H}_{q^*}$ from $\mathcal{H}$ that best fits the working condition the system is currently at, and then  a scheduler receives the index~$q^{*}$ and chooses  the appropriate transmissibility $\mathbf{G}_{q^{*}}$ from $\mathcal{G}$ for calculating the unknown output $y_O$ from the given $\mathbf{y}_{I}$. Note that $\mathcal{H}$ for the Bayes classifier as well as $\mathcal{G}$ for the  scheduler are precalculated during the offline stage.

\subsection{Offline Stage and Online Stage}
\label{sec:offline_estimation}

In this subsection, we elaborate the procedure illustrated in Fig.~\ref{fig:procedure}: 

\begin{enumerate}[(i)]
\item
During the offline stage, two families of transmissibilities~$\mathcal{H}$ and $\mathcal{G}$ are constructed based on offline data sets from different working conditions $\mathcal{C}_1,\dots, \mathcal{C}_q$;
\item 
During the online stage, a Bayes classifier is constructed based on $\mathcal{H}$, and the appropriate transmissibility from $\mathcal{G}$ is assigned to the online data.
\end{enumerate}

\paragraph{Offline Stage}
The transmissibility $\mathbf{H}_q$'s from $\mathcal{H}$ is modeled by the FIR model as follows
\begin{equation}
\label{train FIR1}
y_{I_2, q}(t)=\mathbf{b}_{0, q}^T\mathbf{y}_{I_1, q}(t)+\dots+\mathbf{b}_{n, q}^T\mathbf{y}_{I_1, q}(t-n)+\epsilon_{I, q}(t), 
\end{equation}
where $q=1,\dots,Q$, the sequences $\{\mathbf{y}_{I, q}(t),y_{O, q}(t)\}$ denote the offline data set for $q$-th working condition, $\{\mathbf{b}_{0, q}, \ldots, \mathbf{b}_{n, q}\}$ are parameters, and $\epsilon_{I, q}(t)$ is an i.i.d.~Gaussian random variable with variance $\sigma^2_{I, q}$ given a period of time.

Following Eq.~\eqref{linear regression}, we rewrite Eq.~\eqref{train FIR1} as follows
\begin{equation}
y_{I_2, q}(t)=\mathbf{\phi}_{I_1, q}^T(t) \mathbf{\theta}_{I, q}+\epsilon_{ I, q}(t).
\end{equation}

The estimates of $\mathbf{\theta}_{I, q}$'s and $\sigma^2_{I, q}$'s can be obtained according to Eq.~\eqref{eq:regu_theta} and~\eqref{eq:ridge_variance}, respectively. Note that $\mathbf{G}_q$'s from $\mathcal{G}$ can also be modeled in a similar manner.

\paragraph{Online Stage}

When it comes to assigning the $\mathbf{H}_{q^{*}}$ that best fits the sequences $\{y_{I_2}(t), \mathbf{\phi}_{I_1}(t)\}$, where~$t=n+1,\dots, N+n$, during online estimation, we  assume the sequence  is stationary, and  denote $p(\mathbf{H}_q)$'s as the prior probabilities defined by users for assigning $\mathbf{H}_q$'s. Then, by Bayes formula, we have 
\begin{equation}
\label{eq:full_Bayes} 
p\left(\mathbf{H}_q\mid \mathbf{Y}_{I_2} \right)=\frac{p\bigl(\mathbf{Y}_{I_2}\mid \mathbf{H}_q \bigr)p\bigl(\mathbf{H}_q\bigr)}{p\bigl(\mathbf{Y}_{I_2}\bigr)},
\end{equation}
where~$p\left(\mathbf{H}_q\mid \mathbf{Y}_{I_2} \right)$ is the posterior probability, 
\begin{gather}
\label{eq:likelihood}
  p\bigl(  \mathbf{Y}_{I_2}\mid \mathbf{H}_q\bigr)  =\bigl(2\pi \hat{\sigma}_{I, q}^2 \bigr)^{-N/2}\exp\biggl[-\frac{1}{2 \hat{\sigma}_{I, q}^2}\bigl(\mathbf{Y}_{I_2} - \mathbf{\Phi}_{I_1}\hat{\mathbf{\theta}}_{I, q}\bigr)^2 \biggr] \\
  p\bigl(\mathbf{Y}_{I_2}\bigr)  = \sum_{q =1}^Q p\bigl(\mathbf{Y}_{I_2}\mid \mathbf{H}_q \bigr)p\bigl(\mathbf{H}_q\bigr)\\
  \mathbf{Y}_{I_2}=
\bigl[y_{I_2}(n+1)  \  \cdots \ y_{I_2}(N+n)
\bigr]^T \\
\mathbf{\Phi}_{I_1}=
\bigl[
\mathbf{\phi}_{I_1}(n+1) \  \cdots  \  \mathbf{\phi}_{I_1}(N+n)
\bigr]^T,
\end{gather}
and~$\hat{\mathbf{\theta}}_{I, q}$ and $\hat{\sigma}_{I, q}^2$ are estimates of $\mathbf{\theta}_{I, q}$ and $\sigma_{I, q}^2$ for $\mathbf{H}_q$, respectively. Furthermore,  the best~$\mathbf{H}_{q^{*}}$ from $\mathcal{H}$ is selected as follows
\begin{align}
\label{Bayes log}
\mathbf{H}_{q^{*}}&=\argmax_{\mathbf{H}_q \in \mathcal{H}} p\bigl(\mathbf{H}_q\mid \mathbf{Y}_{I_2} \bigr) \\
&= \argmax_{\mathbf{H}_q \in \mathcal{H}} \log{p\bigl(\mathbf{H}_q\mid \mathbf{Y}_{I_2} \bigr)}\\
&= \argmax_{\mathbf{H}_q \in \mathcal{H}} 
L_q, 
\end{align}
where
\begin{equation}
\label{Lvalue}
L_q = \log{p\bigl(\mathbf{H}_q\bigr)}-N \log{\hat{\sigma}_{I, q}}-\frac{1}{2 \hat{\sigma}_{I, q}^2} \lVert{{\mathbf{Y}_{I_2}-\mathbf{\Phi}_{I_1}\hat{\mathbf{\theta}}_{I, q}}}\rVert_2^2.
\end{equation}

Now according to Eq.~\eqref{linear regression} and Eq.~\eqref{eq:regu_theta}, the estimated output $\hat{y}_O(t)$ can be calculated by
\begin{equation}
\label{estimated target}
\hat{y}_O(t)= \mathbf{\phi}_{I}^T(t)\hat{\mathbf{\theta}}_{O, q^{*}}, \quad t=n+1,\dots, N+n,
\end{equation}
where $\hat{\mathbf{\theta}}_{O, q^{*}}$ is estimated parameter for $\mathbf{G}_{q^{*}}$ obtained during the offline stage.
\begin{remark}
  \label{remark:yi2}
  The choice of~$y_{I_2}$  follows two  principles from  physical insights:
  \begin{enumerate}
  \item
    The Signal-to-Noise Ratios (SNRs)  should be sufficient  when the system   undergoes  different  working conditions;
  \item
    The relation between~$\mathbf{y}_{I_1}$ and~$y_{I_2}$ has to be different under different offline working conditions~$\mathcal{C}_1, \ldots, \mathcal{C}_Q$.
  \end{enumerate}
  Otherwise,~$L_q$'s  in Eq.~\eqref{Lvalue} calculated by~$Q$ auxiliary transmissibilities could be   quite close if the  prior probabilities  are equal, which leads to  ineffectiveness of   the  Bayes classifier. 
\end{remark}

Given the primary-auxiliary model scheduling procedure above, we will see in the next sections that the auxiliary transmissibility family $\mathcal{H}$ which is used for constructing the Bayes classifier is capable of assigning an appropriate transmissibility~$\mathbf{G}_{q^{*}}$ to the online data $\{\mathbf{y}_I(t)\}$. 

\section{Numerical Validation}
\label{sec:numerical-example}
In this section, we present a  numerical example aimed at verifying the effectiveness of the proposed primary-auxiliary model scheduling procedure for signal estimation  problems in switching linear dynamic systems with unknown inputs.

\subsection{Demonstration of  the Primary-Auxiliary Model Scheduling Procedure at Online Stage}
\label{sec:demonstr-model-sched}
\begin{figure}[t!]
\centering
\includegraphics[width=0.5\textwidth,page=1]{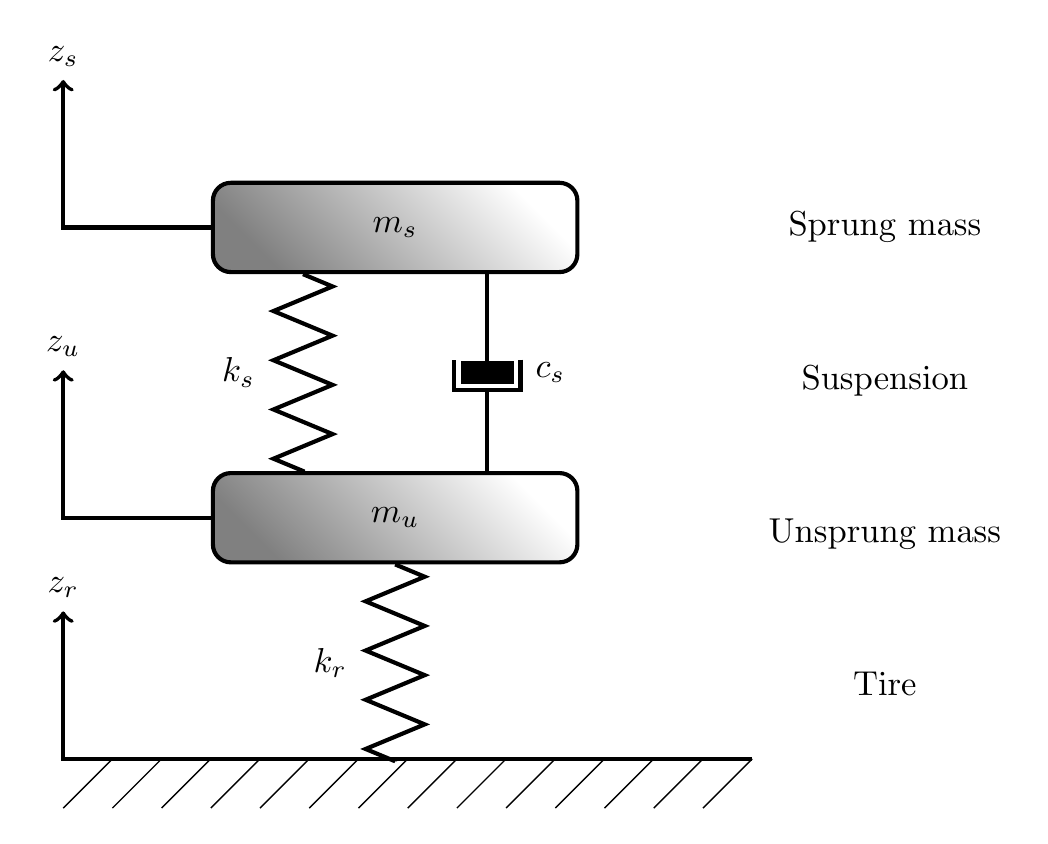}
\caption{A simple quarter-car suspension system.}
\label{fig:suspension0}
\end{figure}

Fig.~\ref{fig:suspension0} shows a quarter-car suspension system, where~$k_r$ is the tire stiffness,~$k_s$ and~$c_s$ are spring and damper coefficients, respectively.    The input of the system  is the displacement   denoted by~$z_r$, and the displacements of the sprung mass and unsprung mass are denoted by $z_s$ and~$z_u$, respectively. Then the state-space form  of the quarter-car suspension system in continuous time   can be represented by
\begin{equation}
  \label{eq:suspension_qtate}
  \begin{aligned}
  \dot{x}(t) = A_c x(t) + b_c z_r(t) \\
  y(t) = C x(t) + d z_r(t),
  \end{aligned}
\end{equation}
where~$x(t)=
  \bigl[
    z_s(t) \
    \dot{z}_s(t) \
    z_u(t) \
    \dot{z}_u(t)
  \bigr]^T
  $,~$y(t) =
\bigl[
     y_{I_1}(t) \  y_{I_2}(t) \  y_{O}(t)
  \bigr]^T
$, and 
\begin{gather}
  A_c  =
  \begin{bmatrix}
    0 & 1 & 0 & 0 \\
    -\frac{k_s}{m_s} &  -\frac{c_s}{m_s} &  \frac{k_s}{m_s} &  \frac{c_s}{m_s} \\
    0 & 0 & 0 & 1 \\
    \frac{k_s}{m_u} &  \frac{c_s}{m_u} &  -\frac{k_s+k_r}{m_u} &  -\frac{c_s}{m_u} 
  \end{bmatrix}, \quad 
  b_c =
  \begin{bmatrix}
    0 \\
    0 \\
    0 \\
    \frac{k_r}{m_u}
  \end{bmatrix}, \\
  C =
  \begin{bmatrix}
    \frac{k_s}{m_u} &  \frac{c_s}{m_u} &  -\frac{k_s+k_r}{m_u} & -\frac{c_s}{m_u} \\
    -\frac{k_s}{m_s} &  -\frac{c_s}{m_s} &  \frac{k_s}{m_s}  &  \frac{c_s}{m_s} \\
    1  &  0  & -1  & 0    
  \end{bmatrix}, \quad
  d =
  \begin{bmatrix}
    \frac{k_r}{m_u} \\
    0 \\
    0
  \end{bmatrix}.
\end{gather}

We would like to estimate the relative displacement~$y_O$  between unsprung mass~$m_u$ and sprung mass~$m_s$ based on~$y_{I_1}$~(the acceleration of~$m_u$)  and~$y_{I_2}$ (the acceleration of~$m_s$). 

In practice, the physical parameters could be different  when  working conditions  change. For sake of simplicity, we consider   two working conditions~$\mathcal{C}_1$ and~$\mathcal{C}_2$, and their physical parameters are listed in Tab.~\ref{tab:physical}.
\begin{table}[htbp!]
  \centering
    \caption{The physical parameters of the quarter-car suspension system with two working conditions.}
  \begin{tabular}[htbp!]{cccc}
    \toprule
    \multirow{2}{*}{Parameters}   &\multirow{2}{*}{Units} &  \multicolumn{2}{c}{Numerical values} \\
  \cline{3-4}
     & & $\mathcal{C}_1$ & $\mathcal{C}_2$ \\
    \midrule 
    $m_s$ &  \si{\kilogram}  & $300$  & $300$\\
    $m_u$ &  \si{\kilogram}  & $40$  & $40$ \\
    $k_s$ & \si{\newton \per \meter} & \num{2.0e4}  & \num{4.0e4} \\
    $k_r$ & \si{\newton \per \meter} & \num{1.8e5}  & \num{2.0e5} \\
    $c_s$ &  \si{\newton \second \per \meter} & \num{1.5e3}  & \num{2.5e3} \\
   \bottomrule
  \end{tabular}
  \label{tab:physical}
\end{table}

The system is discretized using a zero-order hold to obtain
\begin{equation}
  \label{eq:27}
  \begin{aligned}
    x(t+1) &= A x(t) +bz_r(t) \\
    y(t) &= C x(t) +d z_r(t)
    \end{aligned}
\end{equation}
where~$A =  e^{A_c T}$, $b = A_c^{-1}(A-\text{I})b_c$,  and the sampling time~$T$~is~\SI{0.1}{\second}. 
The input is  a realization of a zero-mean, white, Gaussian random process with~\SI{0.01}{\metre\squared} variance.  FIR models with model order~$n =10$ are used to estimate model parameters of the primary and auxiliary transmissibility families.   The ridge regression with~$C_{\text{lim}} =\num{1e6}$ in Section~\ref{regu_MLE_FIR} is applied to tackle ill-conditioning issues in the original linear regression.   The measurements~$y_{I_1}, y_{I_2}, y_O$ are added by zero-mean white Gaussian noise with SNR values of~$50$.  The prior distribution  is~$p\bigl(\mathbf{H}_1\bigr) = p\bigl(\mathbf{H}_2\bigr) = 0.5$.

The training  data set contains~$1000$ data points. The validation data set contains $160$ data points, which is segmented into~$8$ sub-intervals.  Matlab function~\texttt{lsim} is used for plotting simulated time responses of two sub-systems.  

\begin{figure}[t!]
   \centering 
  \includegraphics[width=\textwidth]{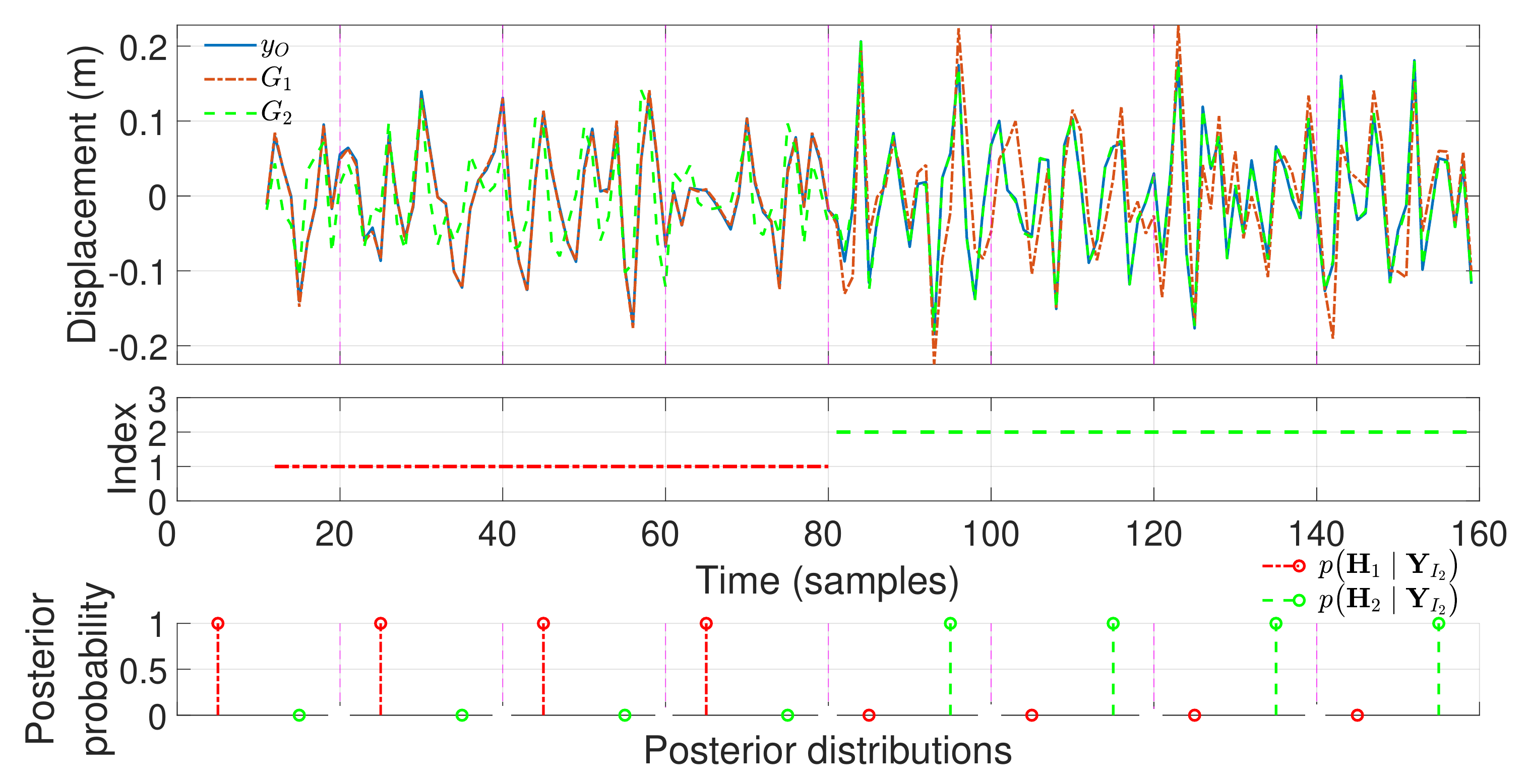}
\caption{The illustration of the  model scheduling procedure. Note that the validation  data set is segmented into~$8$ sub-intervals, where the procedure is sequentially carried out. Top: the target output and its estimates by primary transmissibilities; Middle: evolution of transmissibilities; Bottom: the posterior distributions calculated by Eq.~\eqref{eq:full_Bayes} in Section~\ref{sec:offline_estimation}  for~$8$ sub-intervals. }\label{fig:ill_sch}
\end{figure}

The evolution of switching linear dynamics is as follows: At the beginning, the switching system is running at working condition~$\mathcal{C}_1$  at sample points~$11$--$80$. After~$80$ sample points, the system is switched to working condition~$\mathcal{C}_2$.   Fig.~\ref{fig:ill_sch} shows the illustration of the model scheduling procedure, where the estimation performance of~$y_O$ by primary transmissibilities, the evolution of primary transmissibilities, and  posterior distributions of auxiliary transmissibilities calculated by Eq.~\eqref{eq:full_Bayes} in Section~\ref{sec:offline_estimation} are presented, from which we can observe 
\begin{enumerate}[(i)]
\item
  The primary transmissibility~$G_1$ can successfully estimate~$y_O$ at sample points~$11$--$80$. Correspondingly, the posterior probabilities~$p\left(\mathbf{H}_1\mid \mathbf{Y}_{I_2} \right)$ at sub-intervals~$1$--$4$ are close to~$1$; 
\item
    The primary transmissibility~$G_2$ can successfully estimate~$y_O$ at sample points~$81$--$160$. Correspondingly, the posterior probabilities~$p\left(\mathbf{H}_2\mid \mathbf{Y}_{I_2} \right)$ at sub-intervals~$5$--$8$ are close to~$1$,
  \end{enumerate}
  and it  indicates that the evolution of primary transmissibilities coincide with the evolution of switching linear dynamics. 
  Note that  the transient effect at around data sample~$80$  is fast and can be negligible.
  
Thus,  we can see in this numerical example that FIR models can approximate the transmissibilities very well and the Bayes classifier  can successfully  identify the working condition~$\mathcal{C}_q$.

\subsection{A Comparative Study Between Bayes Classifiers and SVM Classifiers }
\label{sec:svm}
Support Vector Machine (SVM) classifiers are often used for classifying the regressor $\mathbf{\phi}_I(t)$ in Eq.~\eqref{eq:linear_regression_vec} in PWA system identification methods~\cite{garulli2012survey,paoletti2007identification}. The decision rule of the SVM classifier is:

\begin{equation}
\label{eq:SVM_decision_rule}
\begin{aligned}
q^{*}= g(\mathbf{\phi}_I(t)), 
\end{aligned}
\end{equation} 
where $g(\mathbf{\phi}_I(t))$ is a discriminant function which assigns the index~$q^{*}$ to the regressor $\mathbf{\phi}_I(t)$. For a sequence $\{\mathbf{\phi}_I(t), t=n+1,\dots, n+N \}$, it is classified according to a majority vote among the discriminant functions~$ \{g(\mathbf{\phi}_I(t))\mid t=n+1,\ldots, n+N \}$. Matlab function~\texttt{fitcecoc} (the hyperparameters are optimized as far as possible) is used to construct the SVM classifier, which is sequentially carried out in~$12$ sub-intervals, i.e.,~$11$--$20$,~$21$--$40$, $\ldots$,~$221$--$240$,  in validation data.

\begin{figure}[t!]
   \centering 
  \includegraphics[width=\textwidth]{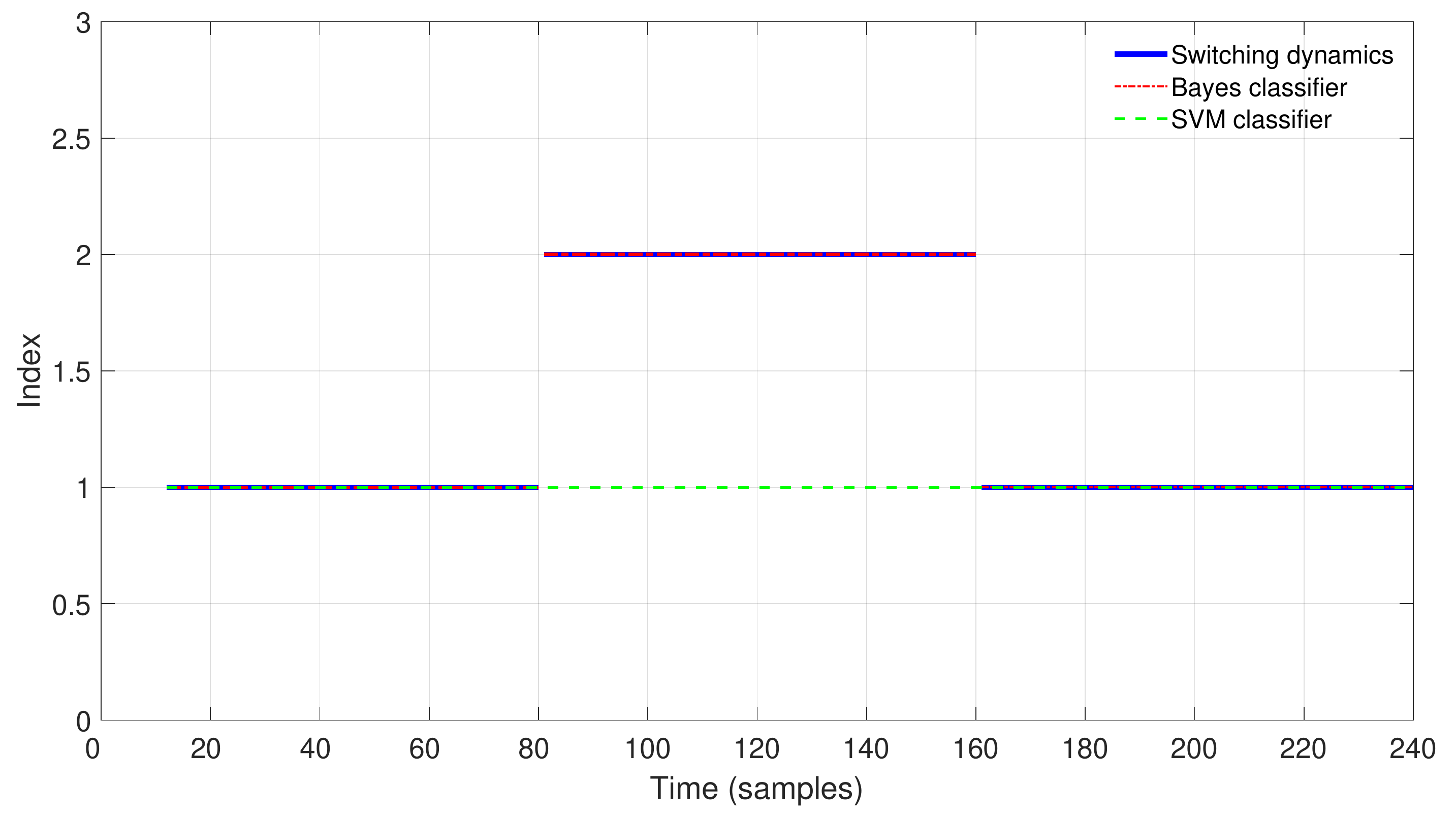}
\caption{The evolution of the switching dynamics as well as  the  transmissibilities from~$\mathbf{y}_I$ to~$y_O$  according to  the Bayes classifier and the SVM classifier.  Note that the validation  data set is segmented into~$12$ sub-intervals (i.e.,~$11$--$20$,~$21$--$40$,~$\ldots$,~$221$--$240$),  where two classifiers are  sequentially carried out.}\label{fig:svm_index}
\end{figure}

Fig.~\ref{fig:svm_index} shows  the evolution of switching dynamics as well as  the  transmissibilities from~$\mathbf{y}_I$ to~$y_O$  according to  the Bayes classifier and the SVM classifier.  In this  example, we can observe the evolution of the  transmissibility obtained from the Bayes classifier coincides with the evolution of switching linear dynamics (at the beginning, the system is running at working condition~$\mathcal{C}_1$, after which it is switched to~$\mathcal{C}_2$, in the end it is switched back to~$\mathcal{C}_1$), while the evolution of the  transmissibility obtained from the SVM classifier  deviates from the evolution of the dynamics  when the system is switched to working condition~$\mathcal{C}_2$. It is not surprising that  the Bayes classifier achieves the excellent  classification accuracy, due to  one-to-one correspondence between  the auxiliary transmissibilities and the  working conditions.     It seems that the SVM classifier cannot  extract the informative  features, and thus fails to identify the working conditions. More discussions on  the success  of Bayes classifiers and  the failure of  SVM classifiers will be given  in  Remark~\ref{remark:success_bayes} and  Remark~\ref{remark:failure_SVM}, respectively, along with the experimental results  in Section~\ref{sec:appl-vehicle-health}.    

\section{Experimental Validation}
\label{sec:appl-vehicle-health}
In this section, we demonstrate the effectiveness of the proposed method by presenting a real-world industrial application to the vertical wheel force  in  a full vehicle system.

\begin{figure}[t!]
  \valign{#\cr
  \hbox{%
    \begin{subfigure}[t]{0.5\textwidth}
      \centering
      \includegraphics[height=5cm,width=0.9\textwidth,page =2]{Figures/suspension_acc}
       \subcaption{The wheel force transducer measuring WCLs   is installed on the wheels. } \label{fig:WFT}
     \end{subfigure}
      }\vfill
  \hbox{%
   \begin{subfigure}[t]{0.5\textwidth}
  \includegraphics[height=5cm,width=0.9\textwidth]{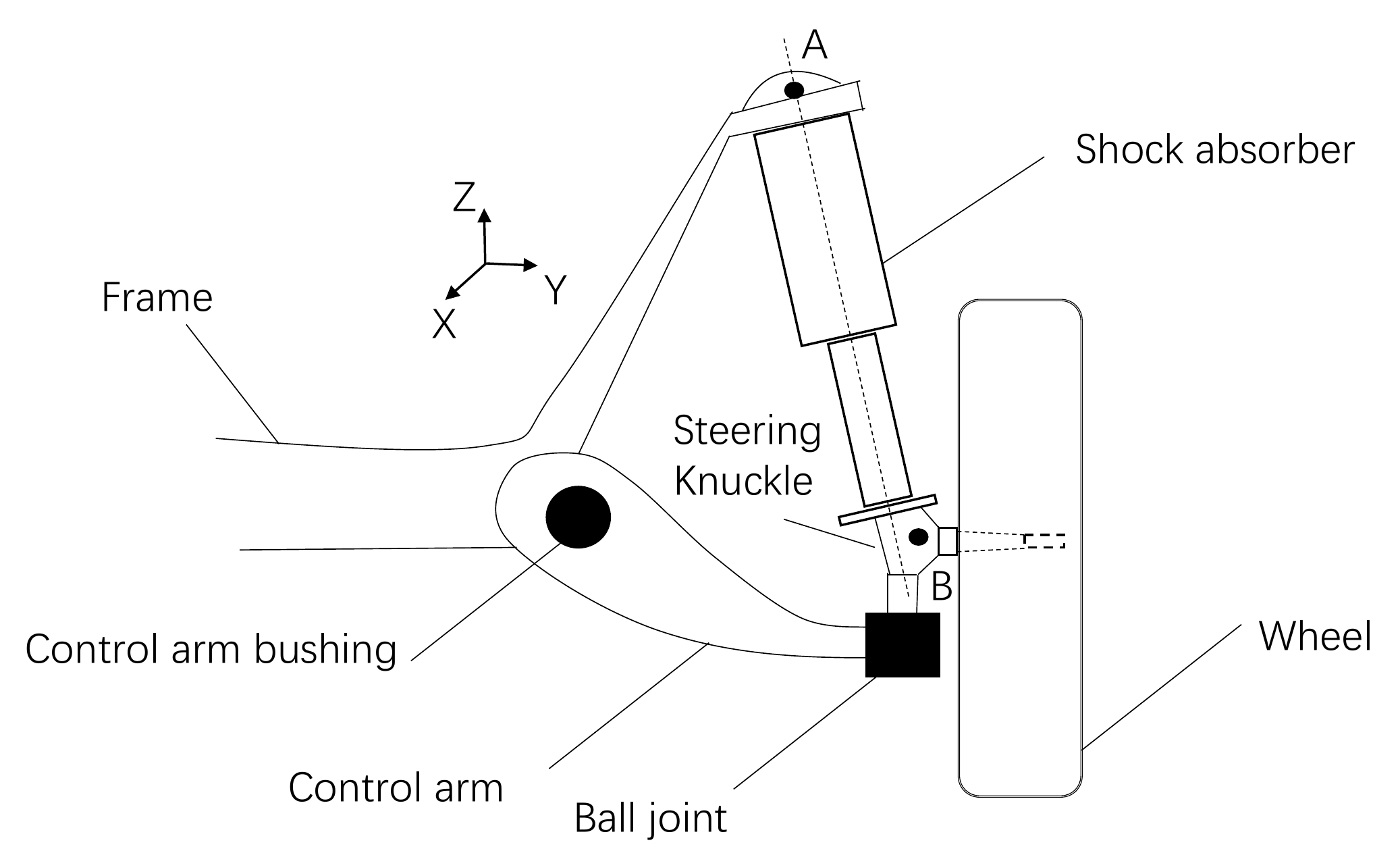}
\subcaption{A sketch of a quarter-car suspension system.} \label{fig:sketch}
\end{subfigure}
}\cr
 \noalign{\hfill}
 \hbox{%
  \begin{subfigure}[t]{0.45\textwidth}
    \centering
    \includegraphics[height=11cm,width=\textwidth,page =1]{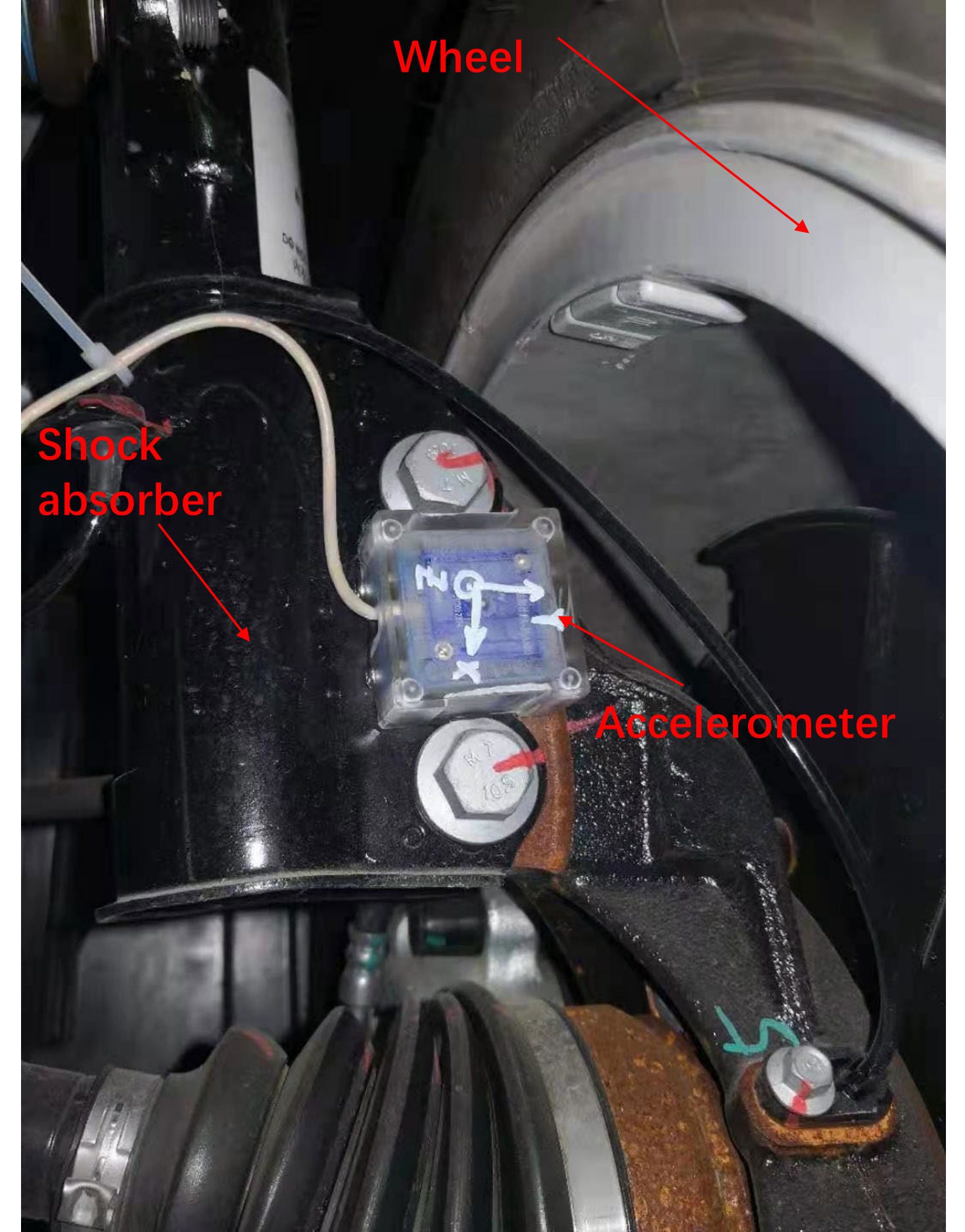}
    \subcaption{The accelerometer is installed  close to  the wheel center.}\label{fig:ACC}
  \end{subfigure}
  }\cr
}
  \caption{The locations of the  accelerometers and the wheel force transducer.}
  \label{fig:location}
\end{figure}

Fig.~\ref{fig:WFT} shows the WFTs measuring WCLs (including the vertical wheel force) in a test vehicle. For the sake of clarity,  Fig.~\ref{fig:sketch} shows the locations of the sensors on a sketch of a quarter-car suspension system: Accelerations in the $X,Y,Z$ directions at the position $A,B$ as well as the suspension deflection  are the known pseudo-inputs~$\mathbf{y}_I$. The vertical wheel force  is  represented by $y_O$, which is the unknown output.   Fig.~\ref{fig:ACC} shows  the accelerometer measuring accelerations at position~$B$ in the full test vehicle. 

\begin{figure}[t]
  \centering
  \begin{subfigure}[t]{0.45\textwidth}
    \centering
    \includegraphics[height=3.5cm, width=0.8\textwidth]{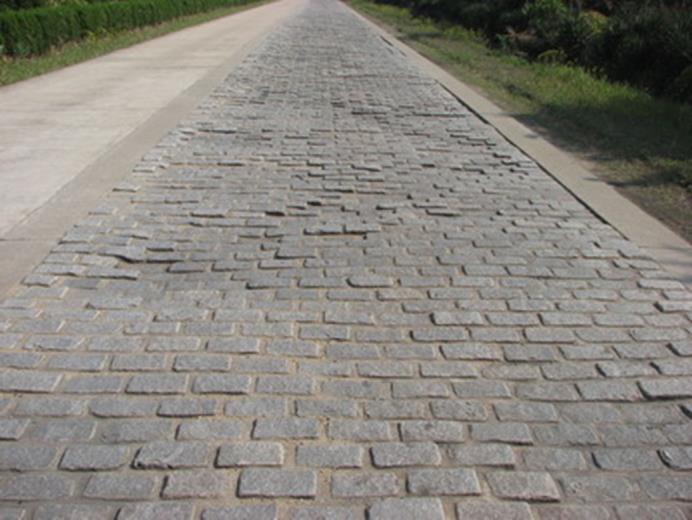}
    \subcaption{Belgium block road profile.}
  \end{subfigure}
\hfill
   \begin{subfigure}[t]{0.45\textwidth}
     \centering
     \includegraphics[height=3.5cm,width=0.8\textwidth]{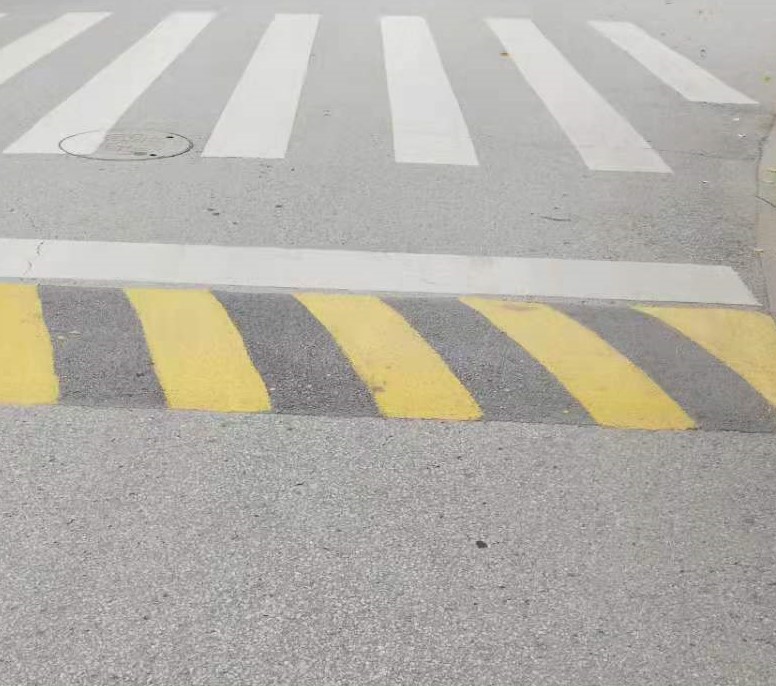}
       \subcaption{Speed bump road profile.}
     \end{subfigure}
  \vfill
   \begin{subfigure}[t]{0.45\textwidth}
     \centering
     \includegraphics[height=3.5cm,width=0.8\textwidth]{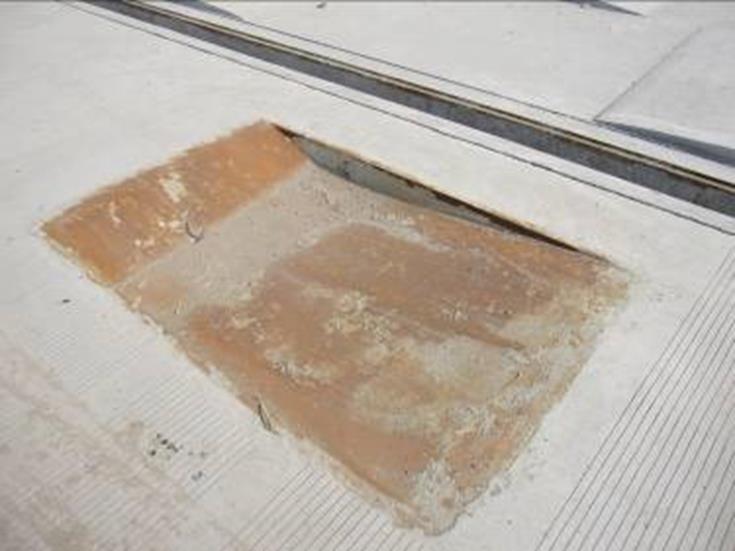}
       \subcaption{Pothole road profile.}
     \end{subfigure}
     \hfill
    \begin{subfigure}[t]{0.45\textwidth}
      \centering
      \includegraphics[height=3.5cm, width=0.8\textwidth]{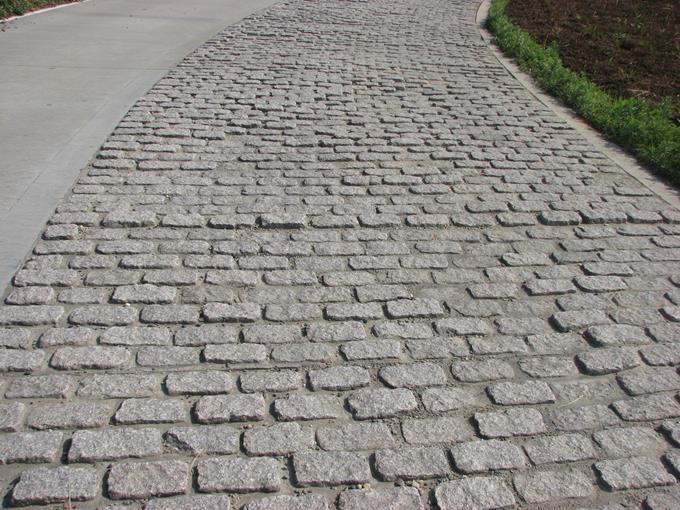}
       \subcaption{Smooth gravel road profile.}
     \end{subfigure}
  \caption{Four  examples of road profiles.}
  \label{fig:road_profile}
\end{figure}

In the experimental runs on a proving ground,~$58$ working conditions (over~\num{2e4} data samples with sampling frequency~\SI{512}{Hz} are collected at each working condition)  are  considered, and they  can be roughly classified into four categories shown in Tab.~\ref{tab:category}.
\begin{table}[t!]
  \small
  \centering
    \caption{Four categories of working conditions on a proving ground}   \label{tab:category}
  \begin{tabular}{cl}
    \toprule
    Category & \multicolumn{1}{c}{Description}  \\ \midrule
    \multirow{2}{*}{1.} & Driving straight on  roads of  random or periodic profiles such as  Belgium \\
             &  block, smooth gravel, and washboard roads  at various vehicle speeds. \\
    &  \\
    2.  & Driving over  bumps, railway crosses, and potholes at various vehicle speeds. \\
    & \\
    3. & Steering  at various vehicle speeds. \\
    & \\
    \multirow{2}{*}{4.}    & Braking on roads of  random or periodic profiles at slight, medium, and\\
                        &   heavy levels.\\
    \bottomrule 
  \end{tabular}
\end{table}
Note that these working conditions cover  the typical ones in RLDA testing campaigns  as many as possible. Fig.~\ref{fig:road_profile}  shows    four  examples of road profiles. The data is preprocessed to be of mean zero.

\subsection{Estimation performances Under Individual Working Conditions}
\label{sec:estim-perf-under}
For a given  working condition, the data set is split into two parts: the training data set and the validation data set. The training data set is used for constructing  an FIR model from~$\mathbf{y}_I$ to~$y_O$ under the given working condition, and the validation data set is used for evaluating  the performance of the constructed FIR model.

The values of the following measure of fit are  introduced~\cite{ljung1995system,bemporad2005bounded}
\begin{equation}
\label{eq:FIT}
\textup{FIT}=100 \cdot \biggl(1-\frac{\lVert{\mathbf{Y}_O-\hat{\mathbf{Y}}_{O}}\rVert_{2}}{\lVert \mathbf{Y}_O-\bar{\mathbf{Y}}_O\rVert_{2}}\biggr)\%, 
\end{equation}
where $\mathbf{Y}_O=\bigl[ y_O(n+1) \ \cdots \  y_O(N+n) \bigr]^T$ are measurements collected from sensors, $\hat{\mathbf{Y}}_{O}$ is the estimate of $\mathbf{Y}_O$, and $\bar{\mathbf{Y}}_O$ is the mean of all entries of $\mathbf{Y}_O$. Eq.~\eqref{eq:FIT} indicates that the performance gets better as FIT gets closer to $100\%$.

\begin{figure}[t!]
  \centering
  \includegraphics[width=\textwidth]{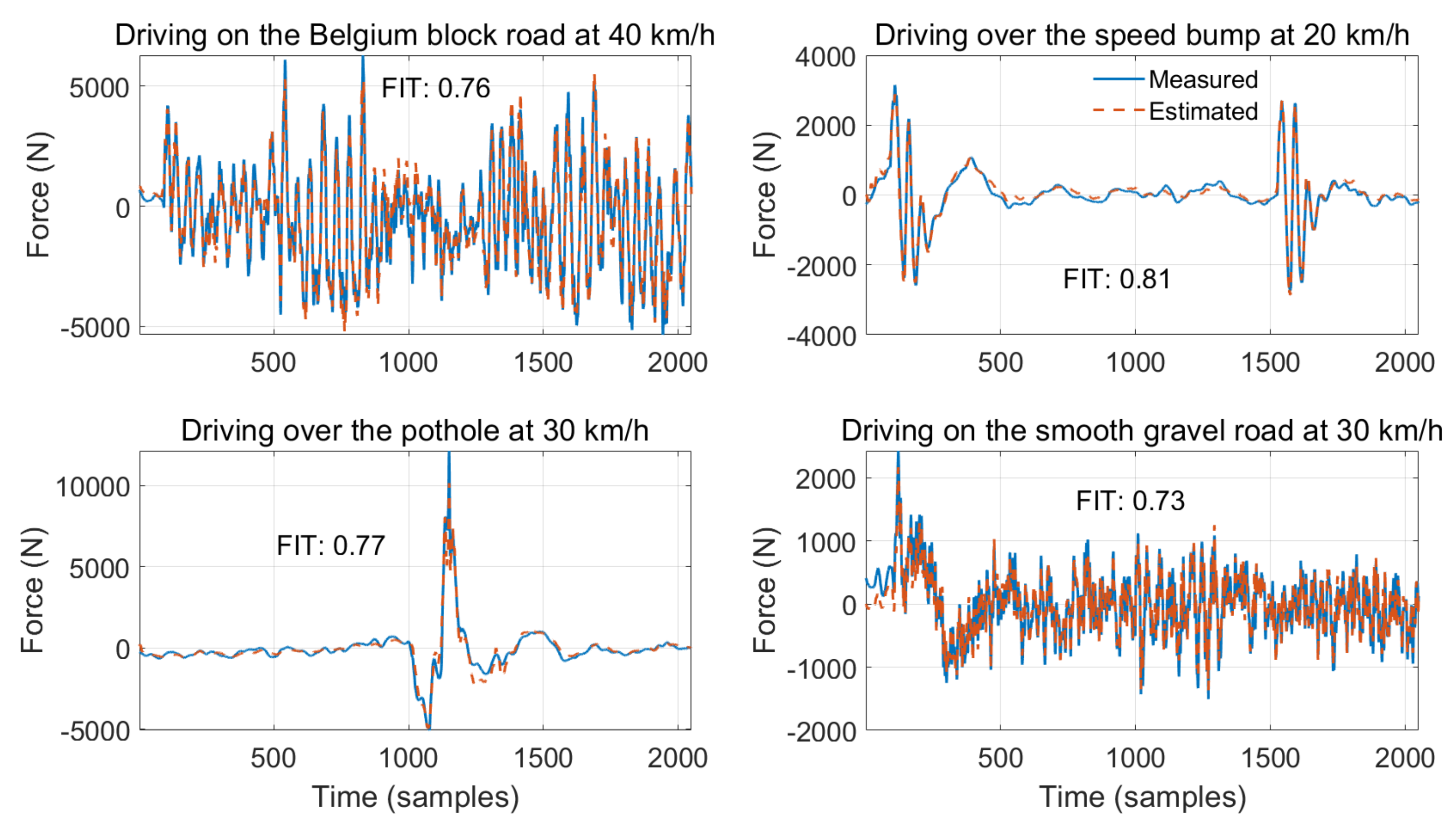}
  \caption{The measurements and estimates when the vehicle is driven on the four  road profiles  shown in Fig.~\ref{fig:road_profile}. The sampling frequency is \SI{512}{Hz}. The data is preprocessed to be of mean zero.}
  \label{fig:profile_val}
\end{figure}

\begin{figure}[t!]
  \centering
  \includegraphics[width=\textwidth]{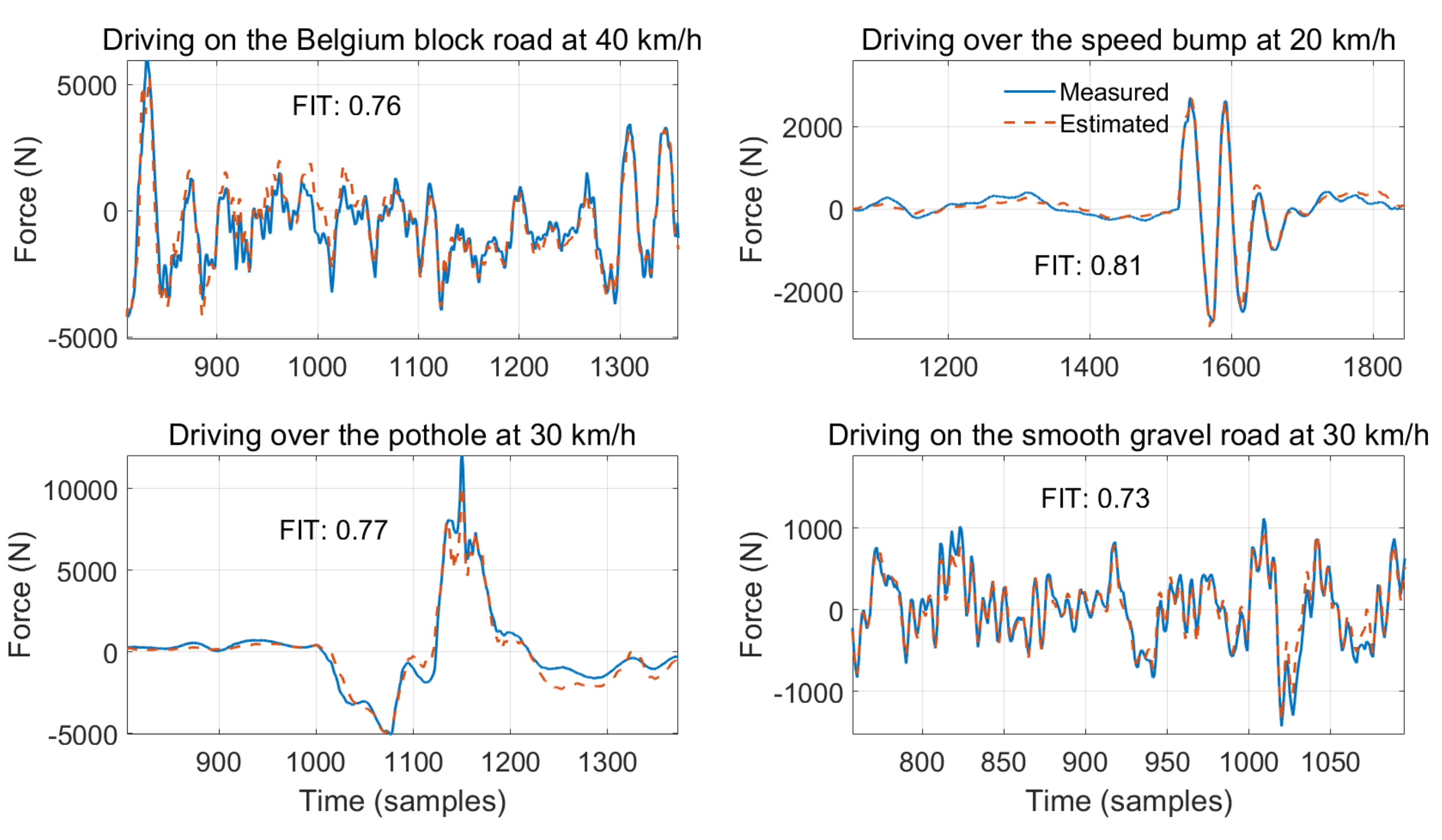}
  \caption{zoom in the result  of Fig.~\ref{fig:profile_val} for the estimation performance  on the four  road  profiles shown in Fig.~\ref{fig:road_profile}. The sampling frequency is \SI{512}{Hz}. The data is preprocessed to be of mean zero.}
  \label{fig:profile_val_zoom}
\end{figure}

Fig.~\ref{fig:profile_val} shows the estimation performances of the vertical force  in validation data sets regarding to four working conditions (i.e., the test vehicle is driven on   four types of road profiles  shown in Fig.~\ref{fig:road_profile}), one by one, and a zoom in the same result is shown in Fig.~\ref{fig:profile_val_zoom}, from which we can observe that good accordance between the measurements and the estimates  is achieved. Furthermore, we consider  the~$58$ working conditions, and  Tab.~\ref{tab:FIT58} shows the good estimation performances  in the sense of FIT, which     validates  that the relation between responses of~$\mathbf{y}_{I_2}$ and~$y_{I_1}$ in  the vehicle system can be approximated by an FIR model  under    a given   working condition.

Next we will show the estimation performances of the proposed primary-auxiliary model scheduling procedure introduced in Section~\ref{sec:patt-recogn-based}.

\begin{table}[t!]
  \centering
    \caption{The estimation performances in the sense of FIT under~$58$ individual working conditions.}\label{tab:FIT58}
  \begin{tabular}{ccccc}
    \toprule 
    mean & standard deviation (std) &    min &     max &     median  \\
    \midrule
80.2\% & 5.6\% & 71.8\% & 90.8\% & 80.7\% \\
    \bottomrule
  \end{tabular}
\end{table}

\subsection{An Illustrative Example of Primary-Auxiliary Model Scheduling Procedure for  Estimation of the Vertical Wheel Force}
\label{sec:illustration}

\begin{figure}[t!]
  \centering
  \includegraphics[width=\textwidth]{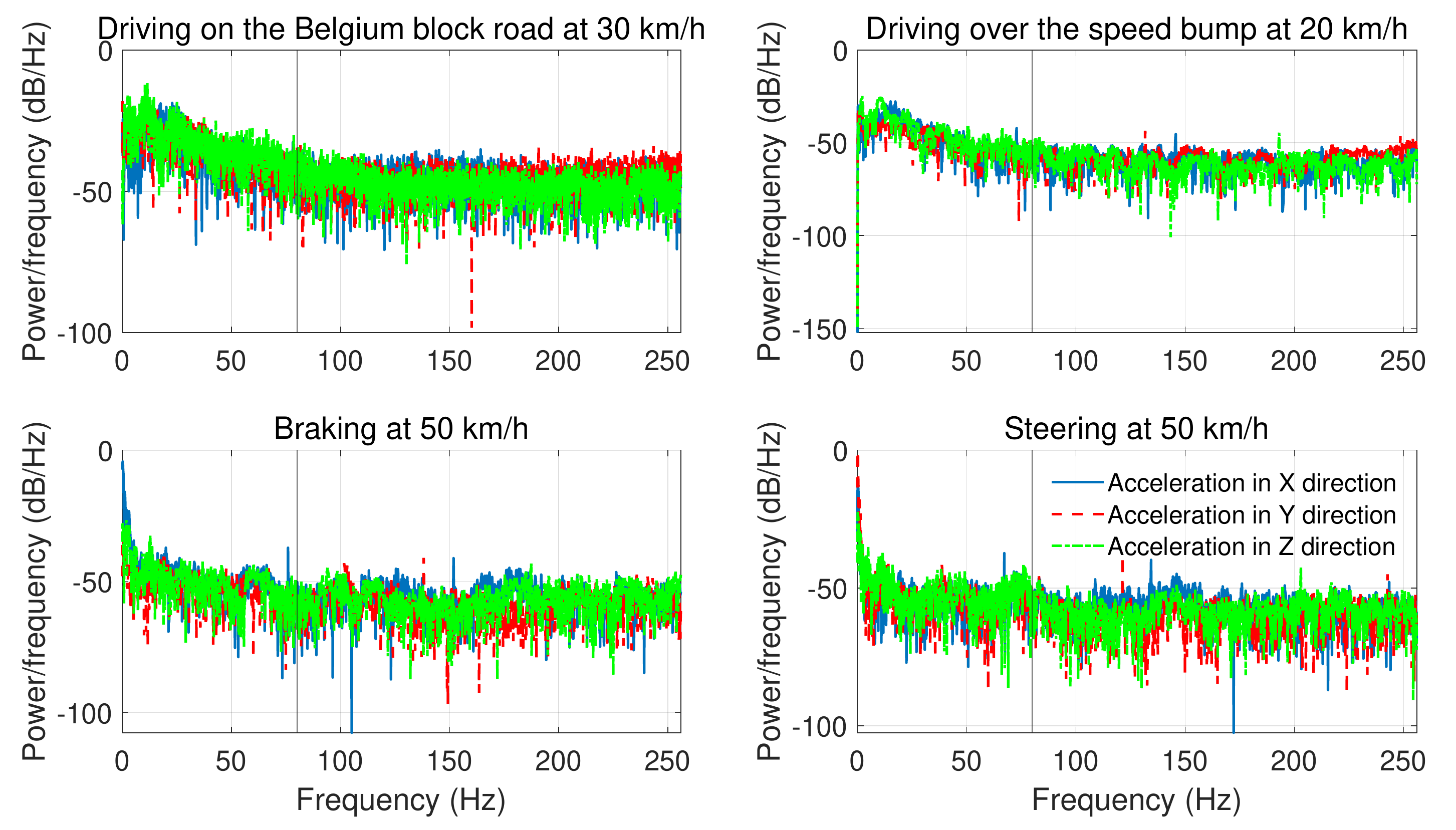}
  \caption{The PSDs of the  accelerations in~$X,Y,Z$ direction at position~$A$ shown in Fig.~\ref{fig:sketch} when the vehicle is driven under four typical working conditions.  It can be observed that SNRs are not sufficient for frequencies above~\SI{80}{Hz}.} 
  \label{fig:acc}
\end{figure}

\begin{figure}[t!]
  \centering
  \includegraphics[width=\textwidth]{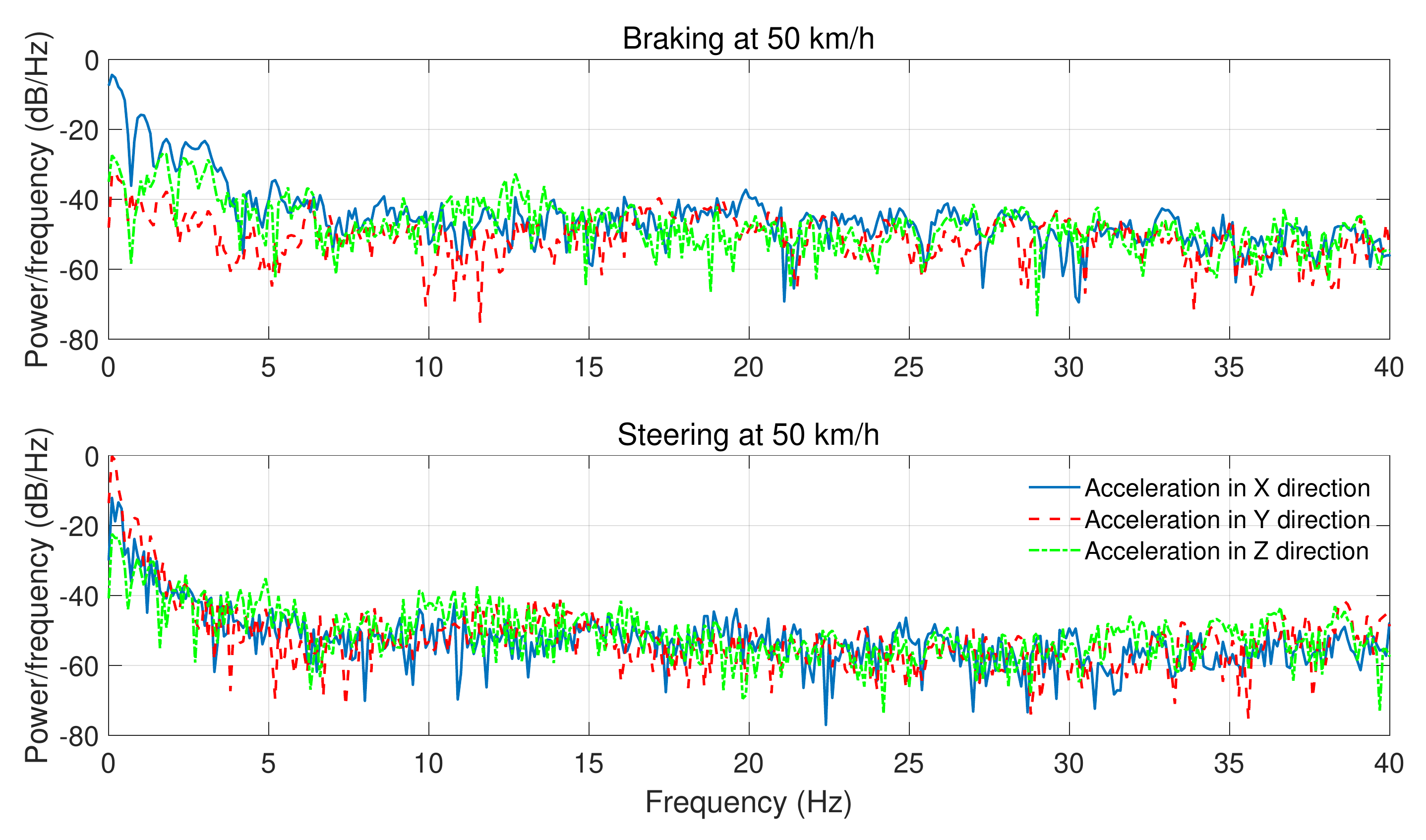}
  \caption{The zoom in the results of the PSDs of the  accelerations in~$X,Y,Z$ direction  under the  braking and  steering working conditions. It can be observed that the SNRs of the acceleration in~$X$ direction  are good at low frequency range from~\SI{0}{Hz} to~\SI{5}{Hz}, while the SNR of the acceleration in~$Y$ direction is poor under the braking working condition.  }. 
  \label{fig:acc_zoom}
\end{figure}

In order to illustrate the primary-auxiliary model scheduling procedure presented in Section~\ref{sec:patt-recogn-based}, five different working conditions are selected from  the four categories shown in Tab.~\ref{tab:category} (two from  category one,  each for one  from remaining three categories) for constructing the transmissibility families $\mathcal{H}$ and $\mathcal{G}$, and another three working conditions are used for the test during the online procedure.   We set the prior distribution as $p\bigl(\mathbf{H}_q\bigr)=\frac{1}{5}$, $k=1,\ldots,5$. 
As mentioned in Section~\ref{sec:patt-recogn-based}, $\mathbf{y}_I$ can be further decomposed into $\mathbf{y}_{I_1}$ and $y_{I_2}$, with $\mathbf{y}_{I_1}(t)\in\mathbb{R}^{n_{I}-1}$ and~$y_{I_2}(t)\in\mathbb{R}$ for each~$t$. Following two  principles in Remark~\ref{remark:yi2}, we denote by $y_{I_2}$ the acceleration in~$X$ direction at  position~$A$ in Fig.~\ref{fig:sketch}, and~$\mathbf{y}_{I_1}$ the remaining accelerations and  the suspension deflection.
\begin{remark}
  From the physical insights, it is easily   understood that the SNR of the acceleration in~$X$ direction is likely to be    sufficient  whenever the driver of the test vehicle is driving  straight on a rough road,  steering, or braking. Furthermore, in order to follow that the relation from~$\mathbf{y}_{I_2}$ to~$y_{I_1}$ has to be different under different offline working conditions (i.e., the second principle in Remark~\ref{remark:yi2} ) as far as possible,  the transmissibility families~$\mathcal{G}$ and~$\mathcal{H}$ are constructed from four different working condition categories shown in Tab.~\ref{tab:category}. 
%
\end{remark}
We may also  verify that the acceleration in~$X$ direction   at position~$A$ in Fig.~\ref{fig:sketch} is suitable for serving as~$y_{I_2}$ in terms of Power Spectral Densities (PSD) signals  in the frequency domain. Fig.~\ref{fig:acc} shows the  PSD  signals of  the accelerations in~$X,Y,Z$ directions when the test vehicle is driven under four working conditions from different categories. For frequencies above~\SI{80}{Hz}, the SNRs of the three accelerations are not sufficient; For frequencies below~\SI{80}{Hz}, the SNRs of the three accelerations are sufficient under the first two working conditions (i.e., the vehicle is driven on Belgium block road or over speed bump). Particularly, a zoom in the same results when the driver of the test vehicle is braking or  steering  is shown in Fig~\ref{fig:acc_zoom}, from which we observe  that the SNRs of the acceleration in $X$ direction are good at the low frequency range from~\SI{0}{Hz} to~\SI{5}{Hz}. On the contrary, the SNR of the acceleration in~$Y$ direction is poor when the driver is braking, which also conforms to  the physical insights. 

\begin{figure*}[t!]
\centering
\includegraphics[width=\textwidth]{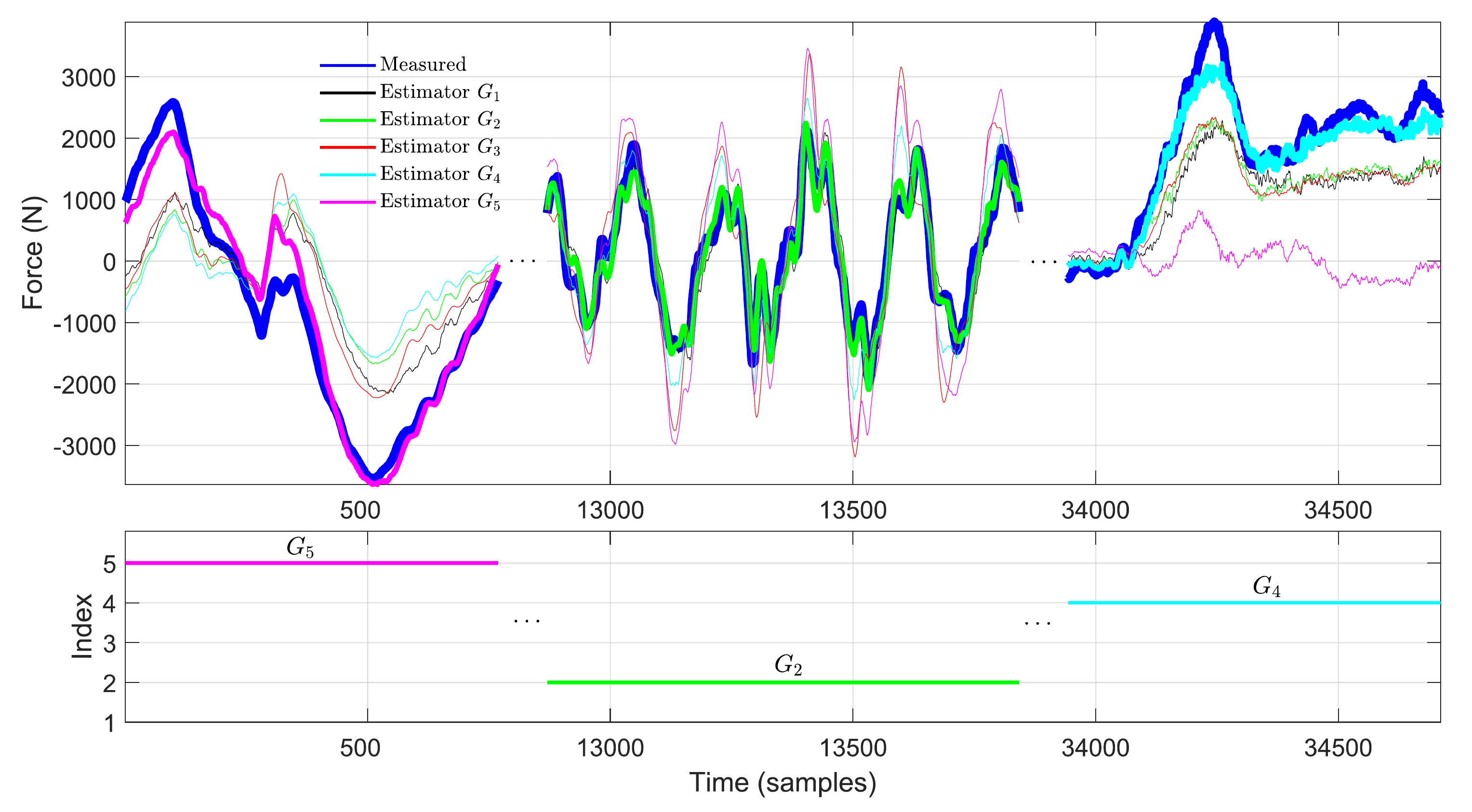}
\caption{The primary-auxiliary model scheduling procedure. Top: the output $y_O$ and estimates from~$\mathcal{G}$. Bottom: evolution of primary transmissibilities. The sampling frequency is \SI{512}{Hz}. The data is preprocessed to be of mean zero.}
\label{fig:evolution}
\end{figure*}

Fig.~\ref{fig:evolution} shows good accordance between the measurement and the estimates from the scheduled estimator obtained by the proposed model scheduling procedure detailed in Section~\ref{sec:patt-recogn-based}, which indicates the Bayes classifier successfully chooses an appropriate estimator under each online working condition. 
\begin{remark}
An online working condition may or may not coincide with one from the offline working conditions for constructing the transmissibility families. Here we choose three online working conditions different from the previous five ones.
\end{remark}

\subsection{Estimation Under Multiple Online Working Conditions}
\label{sec:estim-under-mult}
Furthermore, we consider additional $53$ online working conditions. A comparative study between the scheduled estimator and several other estimators is conducted in Subsection~\ref{sec:individual}, and another  comparative study between Bayes classifiers and a SVM classifier is conducted in Subsection~\ref{sec:svm}.

\subsubsection{A Comparative Study Between Several Estimators}
\label{sec:individual}

Tab.~\ref{tab:working_condition53} shows the performance of the scheduled  estimator, individual estimators,  an ``average'' estimator, as well as  an  ``ideal'' FIT  (as a benchmark) in the aftermath  under $53$ online working conditions in the sense of mean and std of FIT calculated by Eq.~\eqref{eq:FIT}. Note that the individual estimators are $\{\hat{y}_O(t)\}$ calculated from $\mathbf{G}_1,\ldots, \mathbf{G}_5$. The ``average'' estimator is calculated by an ``average'' FIR model whose parameters are obtained by solving a single minimization problem with five offline working conditions putting together, i.e., 

\begin{equation}
\begin{aligned}
\minimize_{\mathbf{\theta}} & & \sum_{q=1}^5\lVert \mathbf{Y}_{O}^q- \mathbf{\Phi}_I^q \mathbf{\theta} \rVert_2^2+\rho \mathbf{\theta}^T \mathbf{\theta}, 
\end{aligned}
\end{equation}
where $\{\mathbf{Y}_{O}^q, \mathbf{\Phi}_I^q\}$ is the data set from offline working condition~$\mathcal{C}_q$. The scheduled  estimator is the one obtained from the model scheduling procedure detailed in Section~\ref{sec:patt-recogn-based}. The ``ideal'' FIT is calculated as 

\begin{equation}
  \label{eq:ideal_FIT}
 \text{FIT}_{\text{Ideal}}(\mathcal{C}_i)=\max_{q\in\{1,\dots,5\}}{\text{FIT}(q;\mathcal{C}_{i})}, \quad i = 1, \ldots, 53, 
\end{equation}
where $\text{FIT}(q; \mathcal{C}_{i})$ denotes the measure of FIT for $q$-th estimator $\mathbf{G}_q$ under~$i$-th online working condition $\mathcal{C}_i$. Tab.~\ref{tab:working_condition53} shows that the scheduled estimator already outperforms the individual estimators and the ``average'' estimator.  Note that  the mean and std of FIT corresponding to the scheduled estimator are close but not identical to those of the ``ideal'' FIT, since the Bayes classifier may sometimes choose a transmissibility from $\mathcal{G}$ that is not optimal for the purpose of estimation (more discussions will be given in  Remark~\ref{remark:ineffecitve}). 
Additionally, the number of offline working conditions for constructing the scheduled estimator is limited,  with more offline working conditions, the performance will be improved in the sense of FIT.

\begin{table}[t!]
\caption{Mean and std of FIT of the vertical force under $53$ working conditions.}
\label{tab:working_condition53}
\centering
\begin{threeparttable}
\begin{tabular}{ccc}
\toprule
Estimators & mean ($\%$) & std ($\%$) \\
\midrule
$\mathbf{G}_1$ & 67.1 & 11.1\\
$\mathbf{G}_2$ &61.9 & 13.2\\
$\mathbf{G}_3$ & 27.2& 29.7\\
$\mathbf{G}_4$ & 64.1& 14.1\\
$\mathbf{G}_5$ & 41.9& 17.7\\
``Average'' estimator & 67.0 & 10.8 \\
Scheduled estimator & 72.2& 10.5 \\
``Ideal'' FIT\tnote{1} & 73.5 & 9.5 \\
\bottomrule
\end{tabular}
\begin{tablenotes}
\item[1] As a benchmark, cf., Eq.~\eqref{eq:ideal_FIT}. 
\end{tablenotes}
\end{threeparttable}
\end{table}

\subsubsection{A Comparative Study Between Bayes classifiers  and SVM classifiers }
\label{sec:svm}


For $i$-th online working condition, where~$ i =1, \ldots, 53$, define an indicator function

\begin{equation}
I_i=
\begin{cases}
 1, & \textup{if } q^{*} \in \argmax_{q\in \mathcal{Q}}{\text{FIT}(q;\mathcal{C}_{i})}, \\
0, & \text{otherwise},
\end{cases}
\end{equation}
where $q^{*}$ denotes the index of the selected $\mathbf{G}_{q^{*}}$ by the Bayes classifier or the SVM classifier.

We also define $\mathcal{A}$ as the classification accuracy in order to compare the performances of the Bayes classifier and the SVM classifier:

\begin{equation}
\label{eq:correctness_rate}
\mathcal{A}=\frac{1}{53}\sum_{i=1}^{53}I_i.
\end{equation}

Tab.~\ref{table:accuracy} shows the classification accuracies of the Bayes classifiers and the SVM classifier, from which we can see that the SVM classifier  (the hyperparameters are optimized as far as possible) has the worst performance of classifying the online working conditions. Also, Tab.~\ref{table:accuracy} shows the result of Bayes classifier with pooled variances. Note that ``pooled variances'' means the variances are identical for all~$\mathbf{H}_q$'s in $\mathcal{H}$, i.e., $\hat{\sigma}^2_{I, q}=\hat{\sigma}^2$ in Eq.~\eqref{Lvalue}. We can see that the information of variances improves the classification accuracy significantly.
\begin{remark}
  \label{remark:success_bayes}
The key reason behind the success of the Bayes classifier in classifying the online working conditions is the strategic construction of the auxiliary transmissibility family $\mathcal{H}$ such that  one-to-one correspondence between~$H_q$ and working condition~$\mathcal{C}_q$ is constructed. Once the working condition is determined, the appropriate primary transmissibility~$G_q$  can be selected to estimate the unknown output. 
\end{remark}
\begin{remark}
  \label{remark:failure_SVM}
The possible reason why the SVM classifier has the worst performance is that the dimension of the regressor domain is much higher than the one in piecewise affine ARX models (In the case study, for a seven input and one output MISO FIR model of order $50$, the dimension of the regressor is $350$)~\cite{garulli2012survey,paoletti2007identification}. Thus, classifying regressors directly with SVM classifiers might be problematic in sensor-to-sensor problems for the purpose of signal estimation, and more complicated  classifiers with nonconvex optimization methods need to be involved, which increases the computational burden and makes the solution hard to assess. 
\end{remark}
\begin{remark}
  \label{remark:ineffecitve}
 The dynamics of the vehicle system  under some  online working conditions is possibly   the same as or is  quite close to the dynamics under corresponding offline working conditions, although the~$53$ online working conditions  differ from the~$5$ offline working conditions. Thus,  the proposed Bayes classifier can choose appropriate primary transmissibilities for estimating the unknown output under these online working conditions (under which the dynamics is the same or is quite close to the one under corresponding offline working conditions). On the contrary,     the proposed Bayes classifier could be ineffective  for    the online working conditions under which the dynamics differs from the one  under~$5$ offline working conditions. Obviously, we can reduce  the limitation by increasing the number of offline working conditions, which also increase the computational cost   in real-time systems. 
\end{remark}
\begin{table}[t!]
\centering
\caption{ The classification accuracies  of three classifiers.}\label{table:accuracy}
\begin{tabular}{cccc}
\toprule
& Bayes & Bayes classifier & SVM \\
& classifier &(pooled variances) & classifier \\
\midrule
Accuracy & $74\%$ & $64\%$ & $41\%$ \\
\bottomrule
\end{tabular} 
\end{table}

\section{Conclusion and Future Works}
\label{sec:conclusion}

In this work,  a novel method is carried out to address  signal estimation problems in nonlinear mechanical systems subject to non-stationary and unknown excitation by constructing pairs of transmissibility families: The system is treated as  a switching linear system, where both the  inputs and the working conditions are  unknown.  In addition to constructing a primary transmissibility family from the pseudo-inputs to the output, an auxiliary transmissibility family is constructed by decomposing the pseudo-input vector into two parts. The auxiliary family constructed from the pseudo-inputs enables to determine  the unknown working condition (which is the bottleneck in the signal estimation problem)  such that an appropriate transmissibility from the primary transmissibility family is selected for estimating the unknown output. 
Furthermore, the   model order  of FIRs, offline working conditions,  and  the outputs of auxiliary transmissibilities are selected in a   sensible  manner. As a result, the proposed approach     offers a   generalizable and explainable    solution to the   signal estimation problems in nonlinear mechanical systems  in the context of  switching linear dynamics  with unknown  inputs.

A real-world application to  the estimation of the vertical wheel force  in a full  test vehicle     is presented to demonstrate the effectiveness of the proposed method. In the illustrative example, good accordance between the measurement and the estimates from the proposed method  is achieved. Also, the proposed method outperforms   the competitive methods in the senses  of FIT and the classification accuracy under multiple  online working conditions.

The following challenge will be addressed in the future work:  the success of proposed model scheduling method  for the estimation of the vertical wheel force  leads to the investigation  of sufficient and necessary conditions under which the best primary transmissibility can be selected correctly according to  the auxiliary transmissibility, which will further  lead to the investigation of a theoretical framework of discrete mode observability~\cite{boukhobza2012sensor} of switching linear systems with unknown inputs in transmissibility contexts. 




 \bibliographystyle{elsarticle-num} 
\bibliography{library}


\end{document}